\documentclass[9pt]{elife}

\usepackage[version=4]{mhchem}
\usepackage{siunitx}
\usepackage{multicol}
\usepackage{enumitem}
\usepackage{booktabs}
\usepackage{array}
\usepackage{graphicx}
\usepackage{amsmath, amsthm}
\usepackage{mathtools}
\usepackage{soul}
\usepackage{natbib}
\usepackage[normalem]{ulem}

\definecolor{structcol}{RGB}{140,40,150}   
\definecolor{dyncol}{RGB}{20,90,190}       
\definecolor{funccol}{RGB}{0,130,110}      
\definecolor{modcol}{RGB}{200,110,0}       

\newcommand{\Scol}[1]{\textcolor{structcol}{#1}}
\newcommand{\Dcol}[1]{\textcolor{dyncol}{#1}}
\newcommand{\Fcol}[1]{\textcolor{funccol}{#1}}
\newcommand{\Mcol}[1]{\textcolor{modcol}{#1}}

\newcommand{\avec}{\mathbf{a}}

\newcommand{\bvec}{\mathbf{b}}

\newcommand{\md}{\mathbf{m}}
\newcommand{\zd}{\mathbf{z}}
\newcommand{\Id}{\mathbf{I}}
\newcommand{\Wd}{\mathbf{W}}

\newcommand{\R}{\mathbb{R}}
\newcommand{\DA}{\mathrm{DA}}
\newcommand{\NA}{\mathrm{NA}}
\newcommand{\ACh}{\mathrm{ACh}}
\newcommand{\HT}{\mathrm{5-HT}}

\title{\huge{Improving the adaptive and continuous learning capabilities of artificial neural networks: Lessons from multi-neuromodulatory dynamics}}

\author[1,2,3,*,\authfn{1},\authfn{3}]{Jie Mei}
\author[4,\authfn{1}]{Alejandro Rodriguez-Garcia}
\author[5]{Daigo Takeuchi}
\author[6,7]{Gabriel Wainstein}
\author[1,8]{Nina Hubig}
\author[9,10]{Yalda Mohsenzadeh}
\author[4,11,*,\authfn{3}]{Srikanth Ramaswamy}
\affil[1]{IT:U Interdisciplinary Transformation University Austria, Linz, Austria}
\affil[2]{International Research Center for Neurointelligence, The University of Tokyo, Tokyo, Japan}
\affil[3]{Department of Anatomy, University of Quebec in Trois-Rivieres, Trois-Rivieres, QC, Canada}
\affil[4]{Neural Circuits Laboratory, Biosciences Institute, Faculty of Medical Sciences, Newcastle University, Newcastle upon Tyne, United Kingdom}
\affil[5]{Graduate School of Medicine, The University of Tokyo, Tokyo, Japan}
\affil[6]{Brain and Mind Centre, The University of Sydney, Sydney, NSW, Australia}
\affil[7]{Helsinki Institute of Life Science (HILIFE), University of Helsinki, Helsinki, Finland}
\affil[8]{Department of Neuroscience, Medical University of South Carolina (MUSC), Charleston, SC, USA}
\affil[9]{Department of Computer Science, Western University, London, Ontario, Canada}
\affil[10]{Vector Institute for Artificial Intelligence, Toronto, Ontario, Canada}
\affil[11]{Theoretical Sciences Visiting Program (TSVP), Okinawa Institute of Science and Technology Graduate University, Okinawa, Japan}

\corr{jie.mei@it-u.at}{}
\corr{srikanth.ramaswamy@newcastle.ac.uk}{}

\contrib[\authfn{1}]{These authors contributed equally to this work}
\contrib[\authfn{3}]{Senior authors}

\begin{document}

\maketitle
\begin{abstract}
Continuous adaptive learning, the ability to adapt to the environment and keep improving performance, is a hallmark of natural intelligence. Biological organisms excel in acquiring, transferring, and retaining knowledge while adapting to volatile environments, making them a source of inspiration for artificial neural networks (ANNs). This study explores how neuromodulation, a building block of learning in biological systems, can help address catastrophic forgetting and enhance the robustness of ANNs in continual learning. Driven by neuromodulators including dopamine (DA), acetylcholine (ACh), serotonin (5-HT) and noradrenaline (NA), neuromodulatory processes in the brain operate at multiple scales, facilitating dynamic responses to environmental changes through mechanisms ranging from local synaptic plasticity to global network-wide adaptability. Importantly, the relationship between neuromodulators and their interplay in modulating sensory and cognitive processes is more complex than previously expected, demonstrating a “many-to-{many}” neuromodulator-to-task mapping. To inspire neuromodulation-aware learning rules, we highlight (i) how multi-neuromodulatory interactions enrich single-neuromodulator-driven learning, (ii) the impact of neuromodulators across multiple spatio-temporal scales, and correspondingly, (iii) strategies for approximating and integrating neuromodulated learning processes in ANNs, and {(iv) an architectural-general formulation of multi-neuromodulatory dynamics.} We {also} present a conceptual study to showcase how neuromodulation-inspired mechanisms, such as DA-driven reward processing and NA-based cognitive flexibility, can enhance ANN performance in a Go/No-Go task. Though multi-scale neuromodulation, we aim to bridge the gap between biological and artificial learning, paving the way for ANNs with greater flexibility, robustness, and adaptability.
\end{abstract}


\section{Introduction}

Learning is the dynamic process by which a system reconfigures itself to improve its performance on a task through experience \citep{kandel_biological_1992, kudithipudi_biological_2022,wang_comprehensive_2024}. In real-world scenarios, the environment is often not stationary, which presents unique challenges in maintaining good performance over time \citep{hadsell_embracing_2020, neftci_reinforcement_2019}. Through evolution, biological organisms have developed the ability to learn a spectrum of tasks over their lifetime with minimal interference. {They} integrate new knowledge without forgetting acquired representations, and {re}use representations shared across tasks, has enabled them to optimize survival \citep{chen_lifelong_2018, flesch_continual_2023, hassabis_neuroscience-inspired_2017, rodriguez-garcia_enhancing_2024, wang_comprehensive_2024}. In recent years, such abilities have been increasingly investigated, and are often considered as a pillar of continual, lifelong learning \citep{chen_lifelong_2018, hadsell_embracing_2020, kudithipudi_biological_2022, wang_comprehensive_2024}.

Despite their success, state-of-the-art artificial neural networks (ANNs) struggle with continual learning \citep{kudithipudi_biological_2022}. In particular, they suffer from catastrophic forgetting when trained sequentially on multiple tasks \citep{mccloskey_catastrophic_1989, goodfellow_empirical_2015}, rely heavily on large labeled datasets, and exhibit limited generalization to out-of-distribution inputs \citep{zador_critique_2019}. {Overall, these modes of failure can be linked to the lack of mechanisms that dynamically regulate when plasticity should be heightened, where in the network it should be expressed, and how it should be implemented in a task-specific manner.} 

{In both the experimental and computational literature, the importance of neuromodulation as a key mechanism underlying knowledge transfer and adaptive behavior has been increasingly acknowledged. However, existing computational studies typically treat the functions of individual neuromodulators in isolation, or conceptualize them at an overly abstract level. The present study is motivated by the need to situate the role of individual neuromodulators within the broader network of neuromodulatory systems. To this end, we characterize the complex interactions between neuromodulators and highlight that their combined effects can extend beyond those attributable to single neuromodulators.}

In this Perspective, we present the dynamics and interactions between multiple neuromodulators (e.g., dopamine, DA; noradrenaline, NA; acetylcholine, ACh; serotonin, 5-HT; histamine, HA) as a mechanism supporting state- and context-dependent control of learning in ANNs. Specifically, we discuss how these neuromodulatory signals individually and collectively regulate plasticity in a spatially and temporally structured manner, and how such synergy contribute to continual learning paradigms. Spatially, individual neuromodulators target different (but commonly overlapping) sub-neuronal compartments, neurons, microcircuits and brain areas. This enables varying rates of learning across the network for flexible reconfiguration of task-specific components and stabilization of shared representations. 

\section{Continual learning in natural and artificial intelligence}
\subsection{Challenges in the development of lifelong learning systems}
{Continual learning imposes computational demands that standard} ANNs trained under supervised or self-supervised paradigms on fixed datasets {often fail to meet \citep{pfeiffer_deep_2018, zador_critique_2019}. Biological nervous systems, on the other hand, satisfy these demands on a daily basis \citep{neftci_reinforcement_2019}. They demonstrate robust learning performance in a noisy and ever-changing environment where information is often partly accessible, carry out a series of actions in response to varying cognitive demands, and integrate new information while significantly preserving existing knowledge, even when it is not relevant to solving the current task at hand. In this section, we outline key areas within the domain of continual learning where artificial systems exhibit misaligned performance with their biological counterparts, and discuss how neuromodulatory systems can inspire novel design principles in each case.}

These models often generalize poorly to \textbf{out-of-distribution (OOD) inputs} and struggle when novel samples or task statistics are present \citep{dangelo_out--distribution_2021, wang_comprehensive_2024}. Achieving high performance often requires large numbers of training examples, implying substantial data storage and computational demands \citep{kudithipudi_biological_2022, schuman_opportunities_2022}. Taken together, these trends reflect an evaluation regime in which AI systems are primarily assessed on their ability to perform inference on static benchmarks, while adaptation, knowledge integration, and learning under distributional shift remain largely outside of the metrics considered.

For this reason, current ANN learning mechanisms based on gradient-based optimization are highly effective in static settings but lead to \textbf{catastrophic forgetting} when tasks are learned sequentially, as parameter updates overwrite representations that were critical for previously learned tasks \citep{mccloskey_catastrophic_1989, goodfellow_empirical_2015}. Importantly, catastrophic forgetting does not arise from limited memory capacity, but from the absence of mechanisms that protect or selectively regulate synaptic plasticity \citep{goodfellow_empirical_2015,kudithipudi_biological_2022}. Furthermore, many proposed methods to address this limitation rely on replay-based strategies \citep{rolnick2019experience}, gradient regularization \citep{kirkpatrick_overcoming_2017, zenke2017continual, aljundi2018memory}, or architectural modularization \citep{parisi_continual_2019, rusu_progressive_2016}, often requiring an oracle signal to indicate task boundaries. This dependence limits their applicability in online, task-agnostic settings characteristic of realistic environments.

This contrast highlights emerging desiderata for artificial systems operating in open-ended settings, where continual knowledge acquisition, integration with prior experience, and adaptive behavior over time are core objectives \citep{hadsell_embracing_2020, kudithipudi_biological_2022, wang_comprehensive_2024}. This has renewed interest in neuroscience, where brain-inspired mechanisms that dynamically regulate learning offer promising routes toward continual adaptation with reduced interference.

\subsection{Continuous, adaptive learning in the biological brain}
To learn in a noisy and ever-changing environment, a series of actions are carried out, with varying cognitive demands \citep{neftci_reinforcement_2019} including: (1) acquiring and tracking knowledge acquired in a completed task, (2) recognizing a new task, (3) evaluating task statistics and similarities between tasks, (4) encoding, reusing and exploiting acquired knowledge, (5) updating and transmitting task-specific variables, and (6) updating internal states during and after learning \citep{hadsell_embracing_2020, kudithipudi_biological_2022}. Through the ability to recognize and exploit shared task structures, the brain can achieve rapid and efficient learning in several trials, a phenomenon commonly depicted in the framework of few-shot learning \citep{kudithipudi_biological_2022, wang_comprehensive_2024}. In other tasks, task shifts do not always occur in a sequence. Accordingly, an organism learns multiple tasks simultaneously, which requires not only storing and retrieving past tasks, but also distinguishing concurrent tasks. In the brain, a continuous learning process is supported by a set of computations:

\begin{itemize}
    \item \textit{\textbf{Encoding task sequences.}} Encoding takes place in  areas such as the prefrontal cortex (PFC), anterior cingulate cortex (ACC) and basal ganglia, where representations of task sequences are stored and retrieved, allowing for flexible execution of complex multi-step behaviors \citep{jin_shaping_2015, takeuchi_cingulate-motor_2022, tanji_role_1994}. Such processes involve the formation of associations between contextual cues, and specific actions and their outcomes, facilitated by synaptic plasticity mechanisms such as long-term potentiation and depression.
    \item \textit{\textbf{Updating task representations.}} As new information comes in, the brain updates its internal models of the task environment by modifying existing representations or creating new ones. Neuromodulators support these processes through signaling the need for behavioral adjustments and triggering synaptic changes necessary to adapt to changing circumstances \citep{akam_what_2021}.
    \item \textit{\textbf{Maintaining task fidelity.}} To ensure stable storage of learned task sequences, the brain needs to avoid catastrophic forgetting. Some regularization methods in ANN exhibit parallels with NMDA-mediated plasticity and clustering of closely related synapses in biological neurons \citep{acharya_dendritic_2022, bono_modeling_2017, kastellakis_linking_2016, limbacher_emergence_2020, pagkalos_leveraging_2024}. However, while they have demonstrated fidelity in supervised image classification tasks, their potential remains unexplored.
\end{itemize}

Given the brain’s ability to learn continuously, catastrophic forgetting, also referred to as \textbf{catastrophic interference}, which takes the form of full erasure of learned representations upon acquisition of new information, is rarely reported in the study of human cognition \citep{french_catastrophic_1999}. Rather, phenomena that are to some extent comparable to catastrophic forgetting, e.g., interference in declarative memory due to overlapping information, have been observed in individuals with amnesia \citep{merhav_neocortical_2014}. Such interference has also been observed in motor learning, where loss of previously acquired motor skills occurs when new skills are learned. When new locomotor tasks are present (e.g., learning a new sport similar to a learned sport but with different rules), there can be interference with existing motor memories, leading to negative transfer of learned motor skills \citep{seidler_neural_2010}. Reduced motor learning and transfer abilities have been shown in brain pathologies: For example, although cerebellar damage does not affect online motor adjustments, it compromises adaptive performance in motor learning processes \citep{morton_cerebellar_2006, smith_intact_2005}. Furthermore, neurodegenerative disorders such as Parkinson's or Huntington's disease can cause decreased performance in kinematic adaptation tasks \citep{laforce_differential_2002}. 

In general, it is difficult to pinpoint cognitive processes that are comparable in nature and functionally analogous to catastrophic forgetting. Nevertheless, studies have focused on identifying the neural correlates of learning and memory, providing a substantial body of evidence on how the nervous system supports learning in a sustained manner \citep{grossman_neuromodulation_2022}.

\section{The neuromodulatory systems}
The brain's neuromodulatory systems play a crucial role in the regulation of neural activity and behavior through the release of chemical substances known as neuromodulators \citep{mei_informing_2022}. Unlike classical neurotransmitters, which act on synapses to enable point-to-point transmission of signals between closely adjacent neurons, neuromodulators also diffuse widely and exert longer-lasting, circuit-level effects, modulating the excitability and plasticity of a group of neurons and consequently contributing to the state and function of the brain \citep{marder_neuromodulation_2012}. This is achieved through mechanisms including adjustment of synaptic strength, modulation of receptor activity, and even alterations in gene expression \citep{marder_neuromodulation_2012, mei_informing_2022}. As a result, neuromodulators do not simply mediate transient communications between neurons, but set the stage for more sustained changes. This capacity to fine-tune brain functions across spatio-temporal resolutions makes them essential for cognitive flexibility and resilience \citep{shine_neuromodulatory_2019}.

\subsection{Key properties}
Neuromodulatory processes can affect neuronal excitability, synaptic plasticity, network dynamics, and ultimately, behavior \citep{marder_neuromodulation_2012, thiele_neuromodulation_2018, nadim_neuromodulation_2014}. They occur across timescales and are responsible for processes ranging from changes in neuronal morphology to alterations in network properties. Thus, the neuromodulatory systems play a key role in brain activity, implicating various cognitive functions, behavior, and emotional states \citep{cools_neuromodulation_2022, thiele_neuromodulation_2018, grossman_neuromodulation_2022}.  

Neurons that produce and release neuromodulators are often clustered in small and well-defined regions of the brain, such as the raphe nuclei (5-HT), the locus coeruleus (NA), the nucleus basalis of Meynert in the basal forebrain (ACh), the substantia nigra (DA) and the tuberomammillary bodies (HA) \citep{mei_informing_2022}. Despite their rather compact origins, these neurons project throughout the brain, pervasively innervating the cortex, thalamus, hippocampus and a multitude of other areas involved in sensory processing, memory and executive functions. The release of neuromodulators helps tune the gain, timing and synchrony of neural circuits, enhancing or dampening the effects of synaptic transmission according to the organism’s physiological state, as well as environmental and behavioral contexts \citep{marder_neuromodulation_2012, thiele_neuromodulation_2018, nadim_neuromodulation_2014}.

The neuromodulatory systems regulate neuronal excitability and synaptic plasticity through complex molecular pathways. For example, NA released from the {locus coeruleus (LC) binds to distinct} G protein-coupled receptor subtypes ($\alpha1$, $\alpha2$, and $\beta$) \citep{benarroch_locus_2018, mcburney-lin_locus_2019}. {Such "one modulator, multiple receptors" binding is characteristic of major neuromodulatory systems and is a source of their functional diversity.} Neuromodulators dynamically adjust neuronal excitability and synaptic plasticity through these diverse pathways, enabling adaptive responses to environmental stimuli.

{A defining feature of neuromodulatory systems, which will be central to the arguments we develop in later sections, is that the mapping between neuromodulators and cognitive processes is neither one-to-one nor unidirectional. Single neuromodulators can engage multiple receptor subtypes with distinct or sometimes opposing effects, whereas a single cognitive function can be shaped by combinations of neuromodulatory drives. Importantly, such complexity is not fully captured by the assumption that each neuromodulator plays a single role, motivating the multi-neuromodulatory perspective we develop in the following sections}.

Understanding how {neuromodulators} together shape neural activity and behavior remains a challenge, requiring data integration across modalities and scales \citep{rodriguez-garcia_enhancing_2024}. Advances in neurotechnology and computational modeling now allow researchers to dissect complex neuromodulatory interactions in unprecedented detail, paving the way for a deeper understanding of how they contribute to the learning capacities of the brain. 

\begin{figure}[t]
\centering
\includegraphics[width=\textwidth]{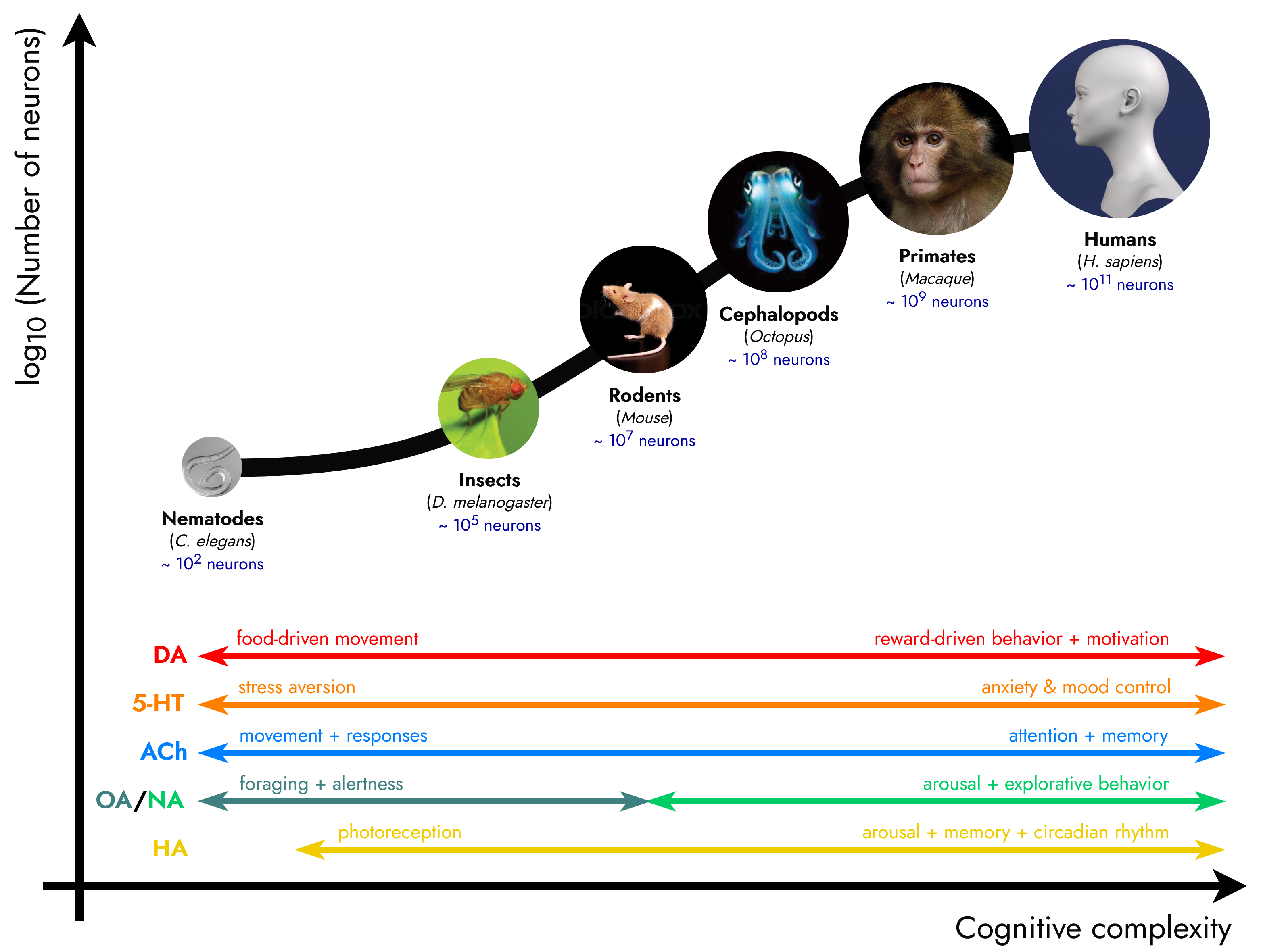}
\caption{\textbf{The role of neuromodulators in species of increasing cognitive complexity.} Although neuromodulatory systems are broadly conserved across species, they exhibit a progressive departure from stereotyped functions as neural and cognitive complexity increases. From nematodes to humans, neuromodulators support a finer functional specification going from basic sensorimotor and homeostatic processes to higher-order functions such as motivation, attention, memory, affect regulation, and adaptive decision making, which in complex brains give rise to social cognition, abstract reasoning, and creative abilities. This expansion illustrates how a shared neuromodulatory basis may enable more advanced and complex behavior across increasing nervous system complexity.}
\label{fig:F1}
\end{figure}

\subsection{Neuromodulatory systems across species}
Given the similar neuromodulatory nuclei and architectures across diverse taxa, neuromodulatory systems are considered to be highly conserved in species spanning from simple invertebrates such as \textit{C. elegans} to birds, reptiles and humans, underscoring their fundamental role in cognition (Figure \ref{fig:F1}). Studying neuromodulatory functions across species provides insights into their evolutionary origins and adaptive significance. Although the complexity of the nervous system varies, the fundamental roles of neuromodulators in perception, cognition and learning highlight their conservation and adaptive value across evolutionary timescales. This comparative approach not only deepens our understanding of fundamental organizational principles of the brain, but also informs research on neurological disorders and the development of targeted therapies.

Throughout evolution, the brain has developed complex and specialized brain regions such as the neocortex, which is responsible for high-order cognitive functions \citep{hamel_milestones_2022, briscoe_evolution_2019,briscoe_homology_2018}. Neuromodulatory nuclei in the brainstem, midbrain, and forebrain form the foundational architecture of the mammalian brain, coordinating broad behavioral and cognitive states through extensive projections \citep{ocallaghan_locus_2021, shine_catecholaminergic_2018, shine_neuromodulatory_2019}. The persistence of neuromodulatory systems underscores their importance in regulating behavior and maintaining neurophysiological balance (Figure \ref{fig:F1}). They enable dynamic adjustments in neural excitability, synaptic plasticity, and network dynamics, which are essential for cognitive flexibility and behavioral adaptability \citep{rodriguez-garcia_enhancing_2024}. {What is conserved across taxa is not a fixed anatomical configuration, but the use of neuromodulatory signals to regulate a broader class of functions encompassing behavioral flexibility, plasticity, attention and arousal, sensory processing, and state- and context-dependent learning.}

\subsection{Neuromodulation in continual learning settings}
Although the connections between experimental and theoretical accounts are still being established, neuromodulatory mechanisms, which impose distinct constraints on how information is acquired, retained, and reused over time, align naturally with the goals of continual learning. Table~\ref{tab:T1} conceptually links continual learning paradigms with relevant experimental frameworks in neuroscience and psychology (see also \citep{avery_neuromodulatory_2017, doya_metalearning_2002, lee_lifelong_2024, montague_computational_2012}). 

\begin{table}[htbp]
\centering

\resizebox{\textwidth}{!}{%
\setlength{\tabcolsep}{6pt}
\renewcommand{\arraystretch}{1.2}
\begin{tabular}{
    >{\raggedright\arraybackslash}p{3.0cm}
    >{\raggedright\arraybackslash}p{6.2cm}
    >{\raggedright\arraybackslash}p{8.2cm}
}
\toprule
\textbf{Paradigm} &
\textbf{Objectives} &
\textbf{Relevant experimental designs} \\
\midrule

\textbf{Transfer learning} &
Efficient adaptation to new domains &
Transfer of learning \citep{haskell_transfer_2006, woodworth_influence_1901} \newline
Mapping of knowledge \citep{gentner_structure-mapping_1983} \newline
Schema learning \citep{tse_schemas_2007} \newline
Structure learning \citep{tervo_behavioral_2014} \\
\midrule

\textbf{Meta-learning} &
Efficient adaptation to new tasks and contexts &
Learning-to-learn \citep{harlow_formation_1949} \newline
Meta-cognitive learning \citep{he2023mechanisms} \\
\midrule

\textbf{Multi-task learning} &
Promoting learning by using shared patterns across tasks and domains &
Dual-task paradigm \citep{pashler_dual-task_1994} \newline
Task switching \citep{rushworth_action_2004, takeuchi_cingulate-motor_2022} \\
\midrule

\textbf{Incremental learning} &
Assimilating new information and efficiently updating models while avoiding catastrophic forgetting &
Retroactive interference \citep{anderson_rethinking_2003} \newline
Task switching \citep{rushworth_action_2004, takeuchi_cingulate-motor_2022} \newline
Set-shifting \citep{dias_dissociation_1996, konishi_transient_1998, robbins_shifting_2007} \\
\midrule

\textbf{Online learning} &
Efficiently updating models upon real-time data collection &
Reversal learning \citep{wilson_orbitofrontal_2014} \newline
Delayed alternation \citep{mishkin_re-examination_1969} \newline
Extinction learning \citep{phelps_extinction_2004} \newline
Set-shifting \citep{dias_dissociation_1996, konishi_transient_1998, robbins_shifting_2007} \\
\bottomrule
\end{tabular}
}
\caption{A summary of continual learning paradigms, their key objectives, and the corresponding experimental paradigms and designs in neuroscience and psychology.}
\label{tab:T1}
\end{table}

The brain is confronted with unique optimization needs and computational demands in each setting. How it enables such an extraordinary feat by orchestrating multiple neuromodulators that act on distinct aspects of learning, and how each neuromodulator functions in each setting, have been addressed by numerous computational and experimental studies.

\textbf{Dopaminergic (DA)} signaling has widely been associated with the reward prediction error signaling \citep{schultz_neural_1997}. Nevertheless, accumulating evidence suggests a broader role in novelty, uncertainty, movement-related variables, as well as the processing of aversive stimuli and threat, revealing DA's involvement in learning and behavioral adaptation \citep{akam_what_2021, avery_neuromodulatory_2017, engelhard_specialized_2019, gershman_explaining_2024, kim_dopamine_2023, lerner_dopamine_2021, matsumoto_two_2009, menegas_dopamine_2018}. Therefore, DA may bias plasticity towards task-relevant representations, supporting memory encoding, and contributing to flexible adaptation in transfer, incremental, and online learning settings \citep{beierholm_dopamine_2013, niv_tonic_2007, lee_dopamine_2021, hattori_meta-reinforcement_2023}.
        
Beyond vigilance, attention, learning, and memory \citep{aston-jones_activity_1981, sara_noradrenergic_1985}, \textbf{noradrenaline (NA)} is relevant to adaptive gain control, network reset, and decision-making under uncertainty \citep{aston-jones_integrative_2005, bouret_network_2005, yu_uncertainty_2005}. These functional roles of NA are linked to the objectives of continual learning paradigms, particularly in transfer, incremental, and online settings, where agents must prioritize task-relevant information and adapt to and encode environmental changes. NA also facilitates multi-task learning, set-shifting, and task-switching by modulating attention and exploratory behavior \citep{shenhav_expected_2013, tervo_behavioral_2014}. Modeling work on the locus coeruleus (LC) suggests that phasic LC activity promotes focused attention, while tonic activity supports behavioral flexibility and exploration \citep{aston-jones_role_1999}. Recent circuit-level studies using optogenetics and GRAB sensors have begun to examine how LC projections to the frontal cortex regulate attention and behavioral switching \citep{bari_differential_2020, su_two_2022}. It is probable that circuits in PFC, ACC, and the striatum differentially and cooperatively process NA signals in continual learning \citep{hassani_noradrenergic_2024}, but the precise mechanisms remain an open question.
    
\textbf{Serotonergic (5-HT)} signaling has been linked to behavioral inhibition and cognitive flexibility \citep{clarke_cognitive_2004}, with computational accounts further relating it to the regulation of temporal discounting and the balance between immediate and delayed rewards \citep{dayan_serotonin_2008, doya_modulators_2008, doya_metalearning_2002}. In continual learning, this can be interpreted as modulating the effective weight assigned to prior knowledge during incremental updates. Moreover, given the role of 5-HT in mood regulation and stress responses \citep{cools_serotonin_2011, dayan_serotonin_2008}, serotonergic dynamics may contribute to managing conflicting task demands and stabilizing task-related parameters in multi-task scenarios.
        
The \textbf{cholinergic (ACh)} system plays a central role in memory encoding, working memory, and attention modulation, partly through its influence on inhibitory interneurons and oscillatory circuit dynamics \citep{hasselmo_role_2006, hasselmo_high_2004}. Such mechanisms support the cross-task integration and stabilization of new information, making ACh particularly relevant for incremental and online learning, where selective pathway modulation may help mitigate interference between competing representations. ACh also facilitates cognitive flexibility under uncertainty by regulating attention and working memory, enabling the prioritization of task-relevant inputs while suppressing distractions in multi-task settings \citep{parikh_prefrontal_2007, sarter_cortical_1999}. In combination with NA, ACh signaling has been proposed to modulate attention allocation during inference in uncertain environments \citep{yu_uncertainty_2005}.
    
The functionality of neuromodulatory systems offers insights for the design of novel ANNs \citep{mei_informing_2022, hassabis_neuroscience-inspired_2017, kudithipudi_biological_2022}, improving learning efficiency through task-appropriate prioritization and and context-aware adjustment of attention \citep{shine_computational_2021}. Integrating neuromodulatory elements may also enhance resilience to disturbances and noise, ensuring stable and robust task performance \citep{shine_neuromodulatory_2019, rodriguez-garcia_enhancing_2024}.

{Across neuromodulators and continual learning paradigms, we observe that no single neuromodulator maps cleanly onto only one function, and no single continual learning paradigm is served by only one neuromodulator. Such many-to-many mapping is a prevalent feature of neuromodulatory dynamics, demanding a multi-neuromodulatory framing in both experimental and computational settings.}

\section{Going beyond single neuromodulators: The multi-neuromodulatory dynamics}
The anatomical signatures of neuromodulator-releasing neurons -- extensive arborization, high density of release sites, and long-range and widespread projections \citep{doucet_quantification_1986, doya_serotonergic_2021, matsuda_single_2009, poe_locus_2020} -- have supported the view of neuromodulation as a spatially diffuse and globally coordinated process associated with brain states \citep{aston-jones_role_1999, aston-jones_integrative_2005, matityahu_acetylcholine_2023}. In parallel, theoretical studies have largely focused on simplified formalizations, and often favor a framework where single neuromodulators are connected to specific cognitive processes (e.g., NA with arousal, DA with reward, 5-HT with cost assessment and ACh with attention), overlooking the effects of their interactions.

{Both framings understate the structural complexity underlying neuromodulatory processes. Neuromodulation is neither uniformly diffuse, nor decomposable into non-interacting processes driven by single neuromodulators. Instead, various types of interactions between neuromodulators have been reported and are more widespread than initially expected. To facilitate the formulation of multi-neuromodulatory dynamics for ANN implementations, here, we distinguish different types of neuromodulatory interactions.}

\subsection{Types of interactions between neuromodulators}

\subsection{Mechanisms underlying multi-neuromodulatory interactions}
In opto- and chemo-genetic studies, one neuromodulator can affect the release and transmission of the other through intricate interconnections \citep{briand_modulators_2007}. Consequently, multiple neuromodulators can together influence a single cognitive task -- spanning from primitive to higher-order functions -- as neuromodulatory receptors participate in processes ranging from sensory perception to complex social and emotional behaviors \citep{froudist-walsh_gradients_2023, hansen_mapping_2022}. Neuromodulatory signaling operates across timescales, continually coordinating the segregation and integration of transient sensorimotor events with longer-term task-level goals \citep{graybiel_habits_2008, shine_catecholaminergic_2018}.

\begin{figure}[ht]
\centering
\includegraphics[width=0.75\textwidth]{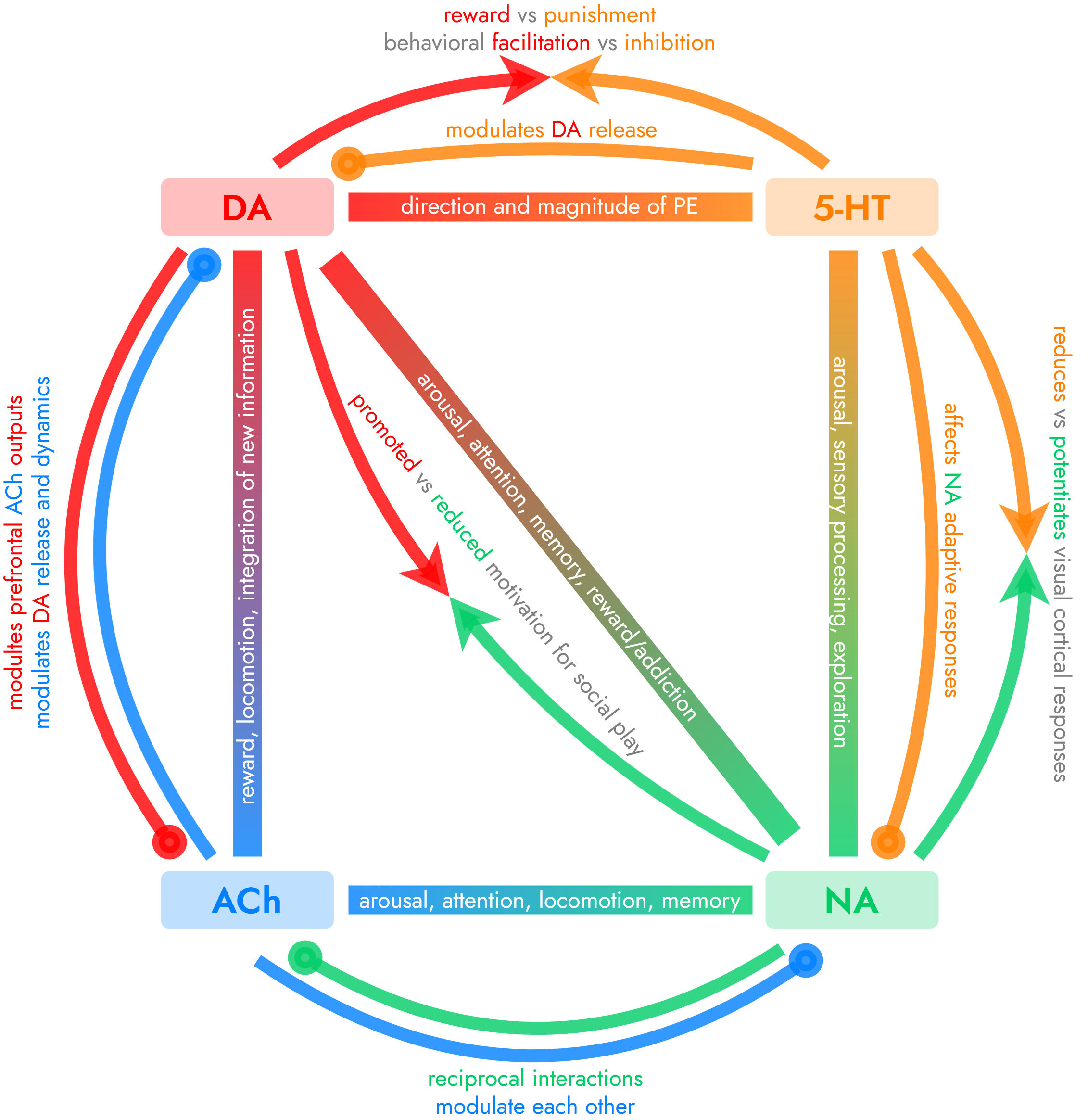}
\caption{\textbf{The complex relationship between neuromodulators.} Modulatory ($\bullet-$): One neuromodulator modulates the release, transmission and/or functional output of the other neuromodulator. Convergent (gradient color bar): Neuromodulators exhibit overlapping, yet sometimes distinctive effects on sensory and cognitive processes. Opponent ($\rightarrow\leftarrow$): Neuromodulators exert opposing effects, or one suppresses the activity of the other. DA: dopamine; 5-HT: serotonin; ACh: acetylcholine; NA: noradrenaline.}
\label{fig:F4}
\end{figure}

The effects of neuromodulation are target-specific: ACh projections from the basal forebrain (BF) differentially shape emotional learning (BF to amygdala), spatial memory (BF to dorsal hippocampus), and cue encoding (BF to medial PFC) depending on their projection sites \citep{likhtik_neuromodulation_2019, zaborszky_specific_2018}. Complementing this region-level specificity, within-region heterogeneity further refines neuromodulatory control in a local range. In the striatum, non-uniform distributions of DA and ACh delineate striosome and matrix compartments with distinct connectivity, neurochemical profiles, and learning-related computations \citep{brimblecombe_striosome_2017, graybiel_striosomes_2023, salinas_dopamine_2016}. These projection-specific mechanisms and within-area heterogeneity collectively demonstrate how neuromodulators can support modular, parallelized processing across areas, while allowing spatially localized processes to exert finer, more transient adjustments.

Neuromodulatory systems have overlapping innervations and their receptors can co-express in the same groups of neurons \citep{grossman_neuromodulation_2022}, enabling spatially structured interactions. Recent research highlights the pervasiveness of parallel operations by neuromodulatory systems across spatio-temporal scales \citep{sippy_unraveling_2023, yagishita_critical_2014}, revealing neuromodulatory interplay at the levels of transmitter dynamics, connectivity properties, and modes of transmission.

\begin{itemize}
   \item \textit{\textbf{Transmitter dynamics.}} Dale’s principle \citep{dale_pharmacology_1935}, later formalized by Eccles as the ‘one neuron, one transmitter’ hypothesis which suggests neurons consistently excite or inhibit \citep{eccles_cholinergic_1954}. Contrary to Dale’s principle, neurons can release two or more neurotransmitters \citep{tritsch_mechanisms_2016, vaaga_dual-transmitter_2014}. Regions that are traditionally considered the source of one particular neuromodulator can release other neuromodulators, e.g., DA release into the dorsal hippocampus by neurons in the LC \citep{kempadoo_dopamine_2016}. The co-release of neuromodulators and neurotransmitters increases the complexity of neuromodulatory functions: In the retina, the spatially non-uniform co-transmission of ACh and gamma-aminobutyric acid (GABA) in starburst amacrine cells allows encoding of direction selectivity in downstream retinal ganglion cells \citep{lee_role_2010}. Such non-uniform co-release is also observed in single neurons, contributing to their context-dependent activities and making them more expressive.
   
    \item \textit{\textbf{Connectivity properties.}} Properties of neural connectivity, i.e., connectivity density, directionality and weighting, may enable cue-dependent switching between multiple perceptual or behavioral strategies. \citep{munn_neuronal_2023} showed that NA and ACh projections contact the same layer V pyramidal neurons through diffuse (NA $\rightarrow$ layer V) and targeted (ACh $\rightarrow$ layer V) innervation patterns, supporting flexibility and reliability respectively. Such concurrent yet differential effects suggest dual-mode information processing within individual neurons and microcircuits. Some areas, such as the striatum, host multiple neuromodulatory systems, allowing the co-modulation of DA, ACh and histamine \citep{cruikshank_dynamical_2023}. Through spatial adjacency as such, neuromodulators not only affect the rate of release of one another to stabilize their concentrations in the extracellular space, but fulfill context-appropriate actions. One example is the coexistence of ACh and DA waves in the dorsal striatum, and that their phase relationship is modulated by the presence of rewards \citep{hamid_wave-like_2021, matityahu_acetylcholine_2023}. These interactions give rise to task- and cue-specific modulation, enhancing the representation of behaviorally relevant information.
    
    Beyond region-level coordination, even a single ACh interneuron can locally regulate striatal DA release, complementing evidence of DA dynamics driven by synchronized ACh activity \citep{matityahu_acetylcholine_2023}. Such interactions can drive fine-tuning and become functionally critical in tasks requiring continuous updates of action-outcome mappings. Disruption of the DA-ACh balance alters the integration of past actions and rewards, leading to inefficient decision switching \citep{chantranupong_dopamine_2023}.
    
     \item \textit{\textbf{Modes of transmission.}} The two primary modes of neuromodulatory transmission are wiring (synaptic) transmission and {volume} transmission \citep{agnati_intercellular_1995, colangelo_cellular_2019, ozcete_mechanisms_2024}. Wiring transmission relies on direct synaptic contacts, enabling spatially precise, energetically efficient, and transient one-to-one signaling. In contrast, volume transmission involves diffusion through the extracellular space, supporting broader one-to-many signaling that is more sustained and less spatially specific. For example, in the ACh system, synaptic transmission mediates precise local modulation, whereas volume transmission conveys sustained, diffuse signals that coordinate broader behavioral states \citep{colangelo_cellular_2019}. {These two transmission} modes {naturally map onto two distinct types of control signal in ANNs: wiring transmission corresponds to highly targeted, precise modulation of subpopulations, whereas volume transmission corresponds to broader, state-dependent coordination that acts on network-level properties such as gain, threshold and activation. With neuromodulation, the two modes are concurrently recruited, combining local specificity with global, network-wide reconfigurations.}
\end{itemize}

\subsection{Multi-neuromodulator dynamics across spatio-temporal scales}


\subsubsection{Subcellular dynamics}
In the nervous system, rapid information transmission is facilitated mainly by ionotropic receptors, which are activated by neurotransmitters such as glutamate \citep{sherman_thalamus_2016} and GABA \citep{staley_ionic_1995}, as well as neuromodulators including 5-HT \citep{barnes_review_1999,thompson_5-ht3_2006} and ACh \citep{albuquerque_mammalian_2009}. These processes drive immediate changes in ion flux through channel opening, influencing short-term synaptic plasticity \citep{hennig_theoretical_2013}.

Metabotropic receptors, which are targeted by most neuromodulators including muscarinic ACh receptors \citep{hasselmo_role_2006}, the majority of 5-HT receptors \citep{barnes_review_1999}, and all receptors for DA \citep{missale_dopamine_1998}, HA \citep{haas_histamine_2008}, and NA \citep{aston-jones_role_1999}, trigger a cascade of second messengers, initiating intracellular biochemical processes \citep{destexhe_synthesis_1994, hasselmo_unexplored_2021}. These relatively slow-acting processes modulate spiking behaviors and enhance long-term synaptic plasticity \citep{hasselmo_unexplored_2021, krichmar_neuromodulatory_2008}.

The coexistence of ionotropic and metabotropic synaptic transmission expands the dimension of the space of neural parameters \citep{hasselmo_unexplored_2021}, enhancing the adaptability of neural networks. The presence of both receptor types in the neuronal membrane allows neural networks to operate across multiple timescales, facilitating continuous learning and ensuring flexibility and resilience in biological systems.

\subsubsection{Neuronal dynamics}
Neurons present specialized input structures called dendrites, which are organized into complex structures known as dendritic trees \citep{branco_single_2010, chavlis_drawing_2021, stuart_dendrites_2016}. Affected by ionic mechanisms that depolarize their membranes, these branched structures propagate diverse non-linear input signals known as dendritic spikes (dSpikes), which attenuate as they travel along the dendrite \citep{acharya_dendritic_2022, pagkalos_leveraging_2024}. 

Biological neural networks exhibit dendritic-spike-dependent plasticity rules that are governed by the timing of synaptic inputs in relation to postsynaptic dendritic spikes rather than axonal action potentials \citep{kampa_dendritic_2007}. These synapses demonstrate scaling as a form of homeostatic plasticity, regulating excitation levels and maintaining the signal-to-noise ratio \citep{rabinowitch_interplay_2006, turrigiano_self-tuning_2008}. Neuromodulators influence dendritic trees by altering their biophysical characteristics in several ways: \textbf{(i)} enhancing and altering ionotropic glutamate or GABA receptors, leading to changes in excitatory postsynaptic potentials (EPSPs) and inhibitory postsynaptic potentials (IPSPs), \textbf{(ii)} releasing Ca2+ to modify the resting potential, and \textbf{(iii)} modifying voltage-gated channels, influencing threshold and refractory period adjustments \citep{shine_computational_2021}. 


\subsubsection{Circuit-level dynamics}

\paragraph{Decision-making} Neuromodulators contribute to decision-making through neural circuits that span the cerebellum, basal ganglia, prefrontal and limbic cortices \citep{cools_neuromodulation_2022,doya_metalearning_2002, grace_dysregulation_2016, grossman_neuromodulation_2022, lisman_hippocampal-vta_2005, sara_locus_2009,schultz_neural_1997}. DA helps reinforce behavior based on reward prediction errors (RPEs) \citep{schultz_neural_1997}, and plays a role in recalibrating the value of actions over time, allowing the brain to adapt to new information \citep{bromberg-martin_dopamine_2010, gardner_rethinking_2018, schultz_updating_2013}. Midbrain DA neurons project to the striatum and prefrontal cortex and modulate synaptic plasticity in these areas, strengthening connections that predict rewarding outcomes and facilitating learning from experiences \citep{doya_metalearning_2002, graybiel_habits_2008, hikosaka_habenula_2010,lisman_hippocampal-vta_2005, watabe-uchida_neural_2017}. Glutamate, the primary excitatory neurotransmitter in the brain, is instrumental in synaptic plasticity and in the formation of neural circuits underlying learning and decision-making. Studies have shown that glutamatergic signaling in the striatum interacts with DA and ACh, and together affect reward-based learning and decision-making \citep{chantranupong_dopamine_2023, krok_intrinsic_2023}. Overall, DA not only reinforces actions based on expected rewards, but contributes to the brain's ability to "re-evaluate" past decisions, which is particularly useful in continual learning \citep{langdon_model-based_2018}.

Accumulating evidence shows that 5-HT also contributes to adaptive decision-making \citep{cardozo_pinto_opponent_2025, cohen_serotonergic_2015, cools_serotonin_2011, daw_model-based_2011}. In \citet{khalighinejad_complementary_2022}, 5-HT and ACh play complementary roles in decision timing and this process involves neural circuits linking the dorsal raphe nucleus (a key source of 5-HT), the basal forebrain (the source of ACh) and the ACC. 

\paragraph{Attention} Neuromodulators control attentional states by dynamically adjusting neuronal circuits to be more receptive to new information or to maintain existing knowledge, depending on task conditions and environmental demands \citep{aston-jones_integrative_2005, bouret_network_2005, hasselmo_role_2006, yu_uncertainty_2005}. ACh adaptively allocates attention for optimized sensory processing \citep{baxter_cognitive_1999, hasselmo_high_2004, sarter_unraveling_2005}, and modulates the activity of cortical neurons, particularly in the prefrontal cortex and sensory areas, to enhance signal-to-noise ratios. It also increases the responsiveness of neurons to relevant sensory inputs while suppressing responses to irrelevant ones \citep{thiele_neuromodulation_2018}, ensuring a focus on task-relevant information. 

DA plays a critical role in adjusting attentional focus based on reward predictions and outcomes. It modulates the allocation of cognitive resources to tasks that are expected to yield high rewards through midbrain DA projections to the prefrontal cortex and basal ganglia \citep{walton2019}. { DA enhances the encoding of reward-related cues and helps prioritize actions that lead to positive outcomes \citep{keiflin2015dopamine, aizenbud2025neural}. This modulation supports the maintenance of motivational states necessary for sustained attention and learning \citep{bromberg-martin_dopamine_2010}.} 
    
In contrast to ACh and DA, NA modulates arousal and stress responses by modulating LC activity \citep{aston-jones_integrative_2005, bouret_network_2005}. \citep{lockhofen_neurochemistry_2021} investigated how NA increases cortical excitability and enhances the detection of salient stimuli by regulating arousal, and suggested that NA helps maintain optimal attentional states that would allow animals to adapt to new information while preserving existing knowledge. The LC/NA system modulates neuronal activity in the PFC, but how attentional control and other cognitive computations such as inhibitory control of behaviors are processed in the LC/NA-PFC circuits remains to be studied \citep{bari_differential_2020, robbins_neuropsychopharmacology_2009}. In summary, neuromodulators collectively facilitate adaptive control of attention and abate the risk of catastrophic forgetting by enhancing task-relevant cues-actions associations, forming robust representations that are resistant to interference.

\paragraph{Memory} Neuromodulators dynamically regulate synaptic plasticity (e.g., LTP and LTD) in circuits in the midbrain, basal ganglia, PFC, entorhinal cortex and the hippocampus, enhancing memory consolidation \citep{fuchsberger_modulation_2022, lammel_reward_2014, lee_dopamine_2021, likhtik_neuromodulation_2019, takeuchi_locus_2016}, strengthening neural representations of previously learned skills. In the PFC, DA signaling  enhances LTP and strengthens connections between contextual cues and task-relevant actions \citep{puig_neural_2015, seamans_principal_2004}. In the amygdala, DA suppresses feedforward inhibition and modulates the time window required for long-term changes to enhance synaptic weights \citep{bissiere_dopamine_2003}. In the dorsal striatum, DA governs long-term changes in both the strengthening and weakening of synaptic connections \citep{pawlak_dopamine_2008}. Apart from DA, NA also promotes synaptic changes such as LTD, which can weaken synapses that are no longer relevant \citep{tully_norepinephrine_2007}. 

Moreover, neuromodulators can selectively “tag” certain synapses for plasticity modulation (i.e., synaptic tagging \citep{frey_synaptic_1997}), ensuring that task-relevant information is preserved while irrelevant connections are pruned \citep{clopath_voltage_2010, moncada_induction_2007, rogerson_synaptic_2014}. \citep{tonegawa_memory_2015} demonstrated a transient increase in hippocampal engram cell excitability following memory reactivation, which  enhances the subsequent retrieval of specific memory contents in response to cues and is reflected in the animal’s precise and effective recognition of contexts. However, how neuromodulators contribute to the formation, maintenance and rapid control of engram cell excitability in neural circuits underlying continual learning and long-term memory that are resistant to catastrophic forgetting remains to be studied.

\section{Introducing multi-neuromodulatory dynamics to ANNs}
\subsection{Relevant deep learning architectures}
Advances in deep learning have significantly expanded the capacity of machine learning and AI, giving rise to diverse architectures that can capture complex data patterns \citep{goodfellow_deep_2016, lecun_deep_2015}, {including convolutional neural networks, recurrent neural networks and spiking neural networks \citep{cox_neural_2014, khaligh-razavi_deep_2014, kubilius2019brain, rajaei_beyond_2019, yamins_using_2016}. Their learning processes are generally governed by globally applied update rules, which constrain task-specific flexibility and plasticity.}

{Among these architectures, spiking neural networks (SNNs) represent the most direct translation of biophysical processes. As an example, their membrane potentials, time constants, thresholds and nonlinear activation functions can be mapped directed to intrinsic parameters of their biological counterparts, providing a platform for testing the effects of metabotropic and ionotropic synaptic transmission.} Moreover, the precise temporal structure of SNNs provides a basis for integrating time-dependent neuromodulatory processes, making them suitable for real-time and online continual learning \citep{eshraghian2023training, ivanov_neuromorphic_2022, schuman_opportunities_2022}.

{More recently, transformer architectures have become a driving force of large-scale models \citep{vaswani2017attention, bommasani_opportunities_2021}. While not  viewed as a neuromodulation-inspired model, they rely on attention mechanisms that capture long-range dependencies and hierarchical structure in sequential data \citep{kozachkov_building_2023, muller_transformers_2024, whittington2022relating}, suggesting a natural locus for neuromodulatory control.} Through self-supervised learning and large-scale optimization, these systems generalize across domains with minimal task-specific fine-tuning \citep{howard2018universal, vaswani2017attention}.


\subsection{Neuromodulation-inspired models in published studies}
Neuromodulation-inspired elements have been implemented in ANNs to achieve flexible learning, robust performance in diverse tasks, and improved adaptation to changing environments. By mimicking the brain's neuromodulatory systems, these models aim to achieve higher levels of computational efficiency and flexibility, similar to what is observed in biological organisms. {However, current implementations rarely capture the highly coordinated, multi-scale dynamics achieved by multiple neuromodulators. Here, we first provide an overview of neuromodulation-inspired models in published studies, and present a formalization that realizes multi-neuromodulatory signaling in a more general framework.}

\subsubsection{Neuromodulation-aware DNNs}
Neuromodulatory-inspired components have been integrated into ANNs through learning rules and hyperparameter adaptation. At a more global level, neuromodulatory signals influence the entire network on slower timescales. These mechanisms are often implemented through context-driven hyperparameter updates. Examples include updating the learning rate and momentum to optimize performance in response to changing conditions \citep{mei_effects_2023, wilson_neuromodulated_2018}, modifying the slope and bias of the activation functions \citep{vecoven_introducing_2020}, or modulating uncertainty to maintain stable learning and prevent catastrophic forgetting \citep{brna2019uncertainty}.



Meanwhile, inspired by how neuromodulatory signals shape the synaptic plasticity window in biological neurons \citep{brzosko_neuromodulation_2019, pedrosa_role_2017, zhang_gain_2009}, studies have incorporated neuromodulation at the connection level, e.g., modulating weight updates through signals such as contextual information \citep{costacurta_structured_2024, daram_exploring_2020, hong_learning_2022, hwu_neural_2020, meshkinnejad2023look, miconi_backpropamine_2020, miconi2023learning, schmidgall_meta-spikepropamine_2023, tang2023neuro, tsuda2026neuromodulators}. These top-down reconfigurations of connectivity can be interpreted as the third factor in the three-factor learning rule $\dot{w} = F(M, \text{pre},\text{post})$ \citep{fremaux_neuromodulated_2016, kusmierz_learning_2017}. Here, $M$ represents the extrinsic, global neuromodulatory signal that guides weight changes in response to environmental changes or shifting task demands, complementing the pre- and post-synaptic activity. In several studies, eligibility traces are used to bridge the gap between fast synaptic events and slower, global neuromodulation, and weight updates may only occur when a modulatory signal is present \citep{barry_fast_2024, liu_cell-typespecific_2021, miconi_backpropamine_2020, schmidgall_meta-spikepropamine_2023}. 

Notably, most modulatory signals considered are related to reward processing or DA signaling \citep{bellec_solution_2020, chung_reinforcement_2020, liu_cell-typespecific_2021, miconi_backpropamine_2020, schmidgall_meta-spikepropamine_2023}. However, recent approaches have expanded the third-factor rule to include surprise signals \citep{barry_fast_2024}, which are more closely associated with the effects of NA. Furthermore, some studies consider multi-neuromodulatory actions, tuning the plasticity window of neuronal connections through combinations like DA and ACh or DA and 5-HT \citep{wert-carvajal_dopamine_2022,zannone_acetylcholine-modulated_2018}, therefore examining opposing and collaborative interactions between neuromodulatory signals. 

{Although} top-down learning signals help address the credit assignment problem in ANNs, {the canonical scalar third-factor abstraction remains structurally limited relative to biological neuromodulation. For example, the neuromodulatory signal \(M\), when treated as a scalar, does not fully capture the opposing, convergent and modulatory interactions (see Figure \ref{fig:F4}). Furthermore, in this formulation, \(M\) is not generally realized as part of the forward computation, leaving out the immediate effects of neuromodulation on intrinsic neuronal properties. Lastly, \(M\) serves as a global signal for weight updates, acting uniformly across the network. Therefore, the spatial specificity of neuromodulatory drives is not represented. More recent studies have attempted to address these structural limitations through cell-cype-specific rules, \citep{liu_action_2022, liu_cell-typespecific_2021}}, or context-aware neuromodulation \citep{costacurta_structured_2024}.

\subsubsection{Computational models of neuromodulation}
To the best of our knowledge, neuromodulation has yet to be implemented at the subcellular and neuronal scale in ANNs. However, studies have attempted to study neuromodulatory processes through theoretical frameworks, investigating the functional roles of neuromodulators in single-cell and network models (for a review, see \citet{fellous_computational_1998}). Fellous and Linster's work probed the activity of neuromodulators through five computational models of progressively increasing biological fidelity: the Markovian kinetics model, the Hodgkin-Huxley model, the FitzHugh-Nagumo model, the leaky integrator model, and the connectionist model. While the biological realism represented in the majority of these models is incompatible with DNNs due to the associated computational costs \citep{rodriguez-garcia_enhancing_2024}, models presented in this review exemplify how neuromodulator dynamics can be flexibly parameterized across abstraction levels. Importantly, the study underscores two significant challenges that still persist today: the absence of direct biological analogs for some neural network parameters and the inability to fully represent all neuromodulation-related processes through parameter changes alone. 

Following this biophysical approach, neuromorphic spiking control systems use fewer neurons with complex dynamics, such as FitzHugh-Nagumo models, to perform motor control tasks in robotics \citep{schmetterling_neuromorphic_2024,sepulchre2022spiking}. These tasks leverage network motifs and neuromodulatory signals to regulate movement through stable and unstable network states \citep{ribar_neuromorphic_2021}. However, scaling these systems to large ANNs is limited by computational costs, posing another challenge in single-cell neuromodulation modeling.

\subsection{{A framework for multi-neuromodulatory dynamics}}

{Many implementations presented above realized neuromodulatory mechanisms at a single spatial scale, or treated the signaling of one specific neuromodulator in isolation from that of other neuromodulators. To offer a complementary view, we characterize multi-neuromodulation as a dynamic yet highly coordinated control layer, and attempt at a unifying formulation that addresses the limitations of a global modulatory signal. For this purpose, our approach highlights how (i) neuromodulatory signals may be implemented across the forward pass, (ii) neuromodulator concentrations evolve and interact, and (iii) intrinsic dynamics, activation functions, and biases of the ANN can be modulated.}

{Instead of creating a new algorithm that accounts for all neuromodulation-relevant details, we propose a formulation of \textit{where} neuromodulation can be incorporated, in \textit{which} forms, and \textit{what} contributions they can make, across spatial and temporal scales. Moreover, this formulation can be connected with published studies, and allows the implementation of new neuromodulation-aware elements building on top of them.}

\paragraph{{Loci of neuromodulator actions}}

{Let $\zd(t) = (z_1(t), z_2(t), \ldots, z_N(t)) \in \R^N$ denote the state vector of a network of $N$ units. The state of unit $i$ can be written as $z_i(t)$, which represents hidden states such as hidden activations or membrane potential, depending on the architecture. Without neuromodulation, the baseline dynamics of a continuous-time model at time $t$ can be expressed as follows:}

{
\begin{equation}
    \boldsymbol{\tau}^{\text{base}} \odot \dot{\zd}(t)=-\,\zd(t)+\Wd^{\text{base}}\,\phi\bigl(\zd(t)\bigr)+\bvec^{\text{base}}+\Id(t),
    \label{eq:base}
\end{equation}
}

{where $\boldsymbol{\tau}^{\text{base}}\in\R_{>0}^{N}$ denotes the unmodulated baseline integration time constants, $\Wd^{\text{base}} \in \R^{N \times N}$ is the baseline connectivity, $\bvec^{\text{base}} \in \R^{N}$ is a constant bias, $\Id(t) \in \R^{N}$ is the external input, and $\phi$ is a pointwise nonlinearity applied to presynaptic states. The homogeneous case used in standard continuous-time RNNs is recovered by setting $\boldsymbol{\tau}^{\text{base}}=\tau^{\text{base}}\mathbf{1}_N$.}

{Let the vector $\md$ describe the instantaneous state of \(K\) neuromodulators. For convenience, we express each component relative to its reference concentration, such that $\md=0$ represents the unmodulated baseline state:}

{
\begin{equation}
    \Mcol{\md(t)}=\big(m_\DA(t),m_\NA(t),m_\ACh(t),m_\HT(t),\ldots,m_K(t)\big)\in \R^K.
    \label{eq:modulators}
\end{equation}
}

{In this section, we identify different loci of neuromodulatory actions and discuss how to realize the effects of neuromodulation in a continuous-time system by making the quantities $\boldsymbol{\tau}$, $\Wd$, $\phi$, and $\bvec$ depend on $\md$.}

{\textbf{Connectivity} $\Wd$ is constructed from a baseline connectivity matrix $\Wd^{\text{base}}$ and is continuously reshaped by neuromodulation. Here, the matrix $\Wd^{\text{base}}$ represents network connectivity at the reference neuromodulatory state. For the $k$-th neuromodulator, where $k \in \{1,\ldots,K\}$, the sensitivity matrix $\Wd^{(k)}$ defines the first-order change in effective coupling associated with neuromodulator $k$. These matrices may be initialized or constrained using biological priors, for example through structured or low-dimensional parameterizations derived from receptor-expression data.}

{
\begin{equation}
    \Scol{\Wd}\bigl(\Mcol{\md(t)}\bigr)=\Scol{\Wd^{\text{base}}}+\sum_k\Mcol{m_k(t)}\Scol{\Wd^{(k)}},\qquad\Scol{\Wd^{(k)}} \in \R^{N \times N}.
    \label{eq:connectivity}
\end{equation}
}

{For example, the $k$-th neuromodulator exerts control over the connection $w_{ij}$ from neuron $j$ to $i$ via $w_{ij}^{(k)}$. At time $t$, the effects of $K$ neuromodulators are pooled through $\sum_k \Mcol{m_k(t)}\,\Scol{w^{(k)}_{ij}}$ and added to the baseline weight $\Scol{w^{\text{base}}_{ij}}$, resulting in the effective weight $\Scol{w_{ij}}\bigl(\Mcol{\md(t)}\bigr)$:}

{
\begin{equation}
    \Scol{w_{ij}}\bigl(\Mcol{\md(t)}\bigr)=\Scol{w^{\text{base}}_{ij}}+\sum_k\Mcol{m_k(t)}\Scol{w^{(k)}_{ij}}.
    \label{eq:connection_ji}
\end{equation}
}

{In addition to the neuromodulator-dependent regulation of connectivity $\Wd$, biological priors could be incorporated into the baseline connectivity $\Wd^{\text{base}}$. Several studies have introduced neuromodulatory actions at this locus \citep{costacurta_structured_2024, bull2025volume}, driven by subnetwork-generated signals or dynamically evolving neuromodulator concentrations.}

{In addition to initializing $\Wd^{\text{base}}$ from a Gaussian distribution, several works have challenged the Gaussian assumption and considered weight distributions that more closely reproduce the heavy-tailed statistics observed in biological connectivity. Specifically, a recent study showed that heavy-tailed RNNs exhibit a broader gain regime near the edge of chaos than Gaussian networks, although at the cost of reduced attractor dimensionality \citep{xie2025slow}. This suggests that the statistics of $\Wd^{\text{base}}$ may interact with neuromodulatory control of neuronal gain.}

{\textbf{Intrinsic parameters} such as integration time constants can also be adjusted in a modulator-dependent manner. In addition, neuromodulators can multiplicatively regulate transmission along individual connections. We therefore define the vector of time constants $\boldsymbol{\tau}$ and a synaptic transmission-gain matrix $\mathbf{G}$ as:}

{
\begin{align}
    \Dcol{\boldsymbol{\tau}}
    \bigl(\Mcol{\md(t)}\bigr)
    &=
    \Dcol{\boldsymbol{\tau}^{\text{base}}}\odot h\!\left(\textstyle\sum_k\Mcol{m_k(t)}\Dcol{\boldsymbol{\rho}^{(k)}}\right),\qquad\boldsymbol{\rho}^{(k)} \in \R^N,
    \label{eq:time_constant}
    \\
    \Dcol{\mathbf{G}}\bigl(\Mcol{\md(t)}\bigr)&=h\!\left(\textstyle\sum_k\Mcol{m_k(t)}\Dcol{\mathbf{K}^{(k)}}\right),\qquad\mathbf{G},\mathbf{K}^{(k)} \in \R^{N\times N}.
    \label{eq:gain_G}
\end{align}
}

{A convenient choice for $h$ is $h=\exp$, applied element-wise, which enforces positivity and allows overlapping modulatory effects to combine multiplicatively. Other positive or bounded mappings could instead be used when saturation or stronger stability constraints are desired. This provides a coarse-grained abstraction of multiplicative interactions that may arise through metabotropic signaling. Here, the integration time constant $\tau_i$ sets the memory horizon of unit $i$.}

{The coupling gain between units $j$ and $i$, denoted by $G_{ij}$, dynamically rescales synaptic transmission along that connection. Thus, Equations \eqref{eq:connectivity} and \eqref{eq:gain_G} represent complementary forms of synaptic modulation: $\Wd(\md)$ captures additive reconfiguration relative to the baseline connectivity, whereas $\mathbf{G}(\md)$ captures multiplicative regulation of transmission efficacy. Individual implementations need not employ both mechanisms simultaneously. Similar to the sensitivity matrix $\Wd^{(k)}$, the receptor-sensitivity parameters $\boldsymbol{\rho}^{(k)}$ and $\mathbf{K}^{(k)}$ can be initialized or constrained according to receptor-expression data for the $k$-th neuromodulator, allowing cell- and pathway-level specificity.}

{Under these definitions, neuronal integration timescales and transmission gains are dynamically tuned by neuromodulator concentrations \citep{shine_neuromodulatory_2019}. For example, concurrent regulation of network integration timescales in spinal motor systems by 5-HT and NA \citep{Heckman2008, Thorstensen2024} can be formulated as:}

{
\begin{align}
    \Dcol{\tau_i}\Mcol{(\md(t))}&=\Dcol{\tau_i^\text{base}}\exp\!\left(\Dcol{\rho_i^{(\HT)}}\Mcol{m_\HT(t)}+\Dcol{\rho_i^{(\NA)}}\Mcol{m_\NA(t)}\right),
    \label{eq:tau}
    \\
    \Dcol{G_{ij}}\Mcol{(\md(t))}&=\exp\!\left(\Dcol{K_{ij}^{(\HT)}}\Mcol{m_\HT(t)}+\Dcol{K_{ij}^{(\NA)}}\Mcol{m_\NA(t)}\right).
    \label{eq:g}
\end{align}
}

{Beyond the two above-mentioned loci, neuromodulators also adjust how each unit \textbf{transforms its state into an output}. At this level, neuromodulation can act on the neuronal gain $\avec$ and the activation threshold $\boldsymbol{\theta}$:}

{
\begin{align}
    \Fcol{\avec}(\Mcol{\md(t)})&=h\!\left(\textstyle\sum_k\Mcol{m_k(t)}\Fcol{\boldsymbol{\mu}^{(k)}}\right),\qquad\Fcol{\boldsymbol{\mu}^{(k)}}\in\R^{N};
    \label{eq:gain_a}
    \\
    \Fcol{\Delta\boldsymbol{\theta}}(\Mcol{\md(t)})&=\textstyle\sum_k\Mcol{m_k(t)}\Fcol{\boldsymbol{\nu}^{(k)}},\qquad\Fcol{\boldsymbol{\nu}^{(k)}}\in\R^{N}.
    \label{eq:threshold}
\end{align}
}

{For unit $i$, the gain $a_i$ rescales the slope of the transfer function, thereby modulating the unit's sensitivity to its input. The shift in activation threshold $\Delta\theta_i$ represents the extent to which neuromodulation changes the state required before unit $i$ responds. More generally, for activation functions without a strict threshold, $\Delta\theta_i$ can be interpreted as a shift in the operating point of the unit. We restrict $a_i$ to positive values because a negative gain would reverse the monotonicity of the unit's input-output transformation rather than simply rescaling its sensitivity.}

{The vectors $\boldsymbol{\mu}^{(k)}$ and $\boldsymbol{\nu}^{(k)}$ denote the modulator-specific gain and threshold profiles, specifying how neuromodulator $k$ selectively scales unit sensitivity and shifts response thresholds. To keep these two effects distinct, gain acts on the state relative to the shifted threshold, such that the modulated state-to-output transformation is}

{
\begin{equation}
    \mathbf{r}(t)=\phi\!\left(\Fcol{\avec}(\Mcol{\md(t)})\odot\left[\zd(t)-\Fcol{\Delta\boldsymbol{\theta}}(\Mcol{\md(t)})\right]\right),
    \label{eq:modulated_output}
\end{equation}
}

{where $\mathbf{r}(t)\in\R^N$ denotes the instantaneous output of the units. This parameterization prevents changes in neuronal gain from implicitly changing the location of the activation threshold.}

{In addition, neuromodulation can introduce more sustained changes in the baseline state of the system by modifying the bias $\bvec^{\text{base}}$, reflecting, for example, state-dependent changes in excitability associated with arousal or behavioral context. The resulting additive drive can be defined as:}

{
\begin{align}
    \Fcol{\bvec}(\Mcol{\md(t)})
    &=
    \Fcol{\bvec^{\text{base}}}+\textstyle\sum_k\Mcol{m_k(t)}\Fcol{\boldsymbol{\beta}^{(k)}},
    \qquad \Fcol{\boldsymbol{\beta}^{(k)}}\in\R^{N}.
    \label{eq:bias}
\end{align}
}

{The additive-drive sensitivity vector $\boldsymbol{\beta}^{(k)}$ controls how neuromodulator $k$ modifies unit baseline excitability. For example, for a neuromodulatory signal associated with increased arousal, $\beta_i^{(k)}>0$ would provide an increasingly depolarizing drive to unit $i$. When the corresponding neuromodulatory state varies slowly, this term can approximate a tonic change in network excitability.}

{Substituting \eqref{eq:connectivity}, \eqref{eq:time_constant}, \eqref{eq:gain_G}, and \eqref{eq:gain_a}--\eqref{eq:bias} into \eqref{eq:base} gives the dynamics of the neuromodulated network:}

{
\begin{equation}
    \Dcol{\boldsymbol{\tau}}(\Mcol{\md(t)})\odot\dot{\zd}(t)=-\zd(t)+\Bigl[\Scol{\Wd}(\Mcol{\md(t)})\odot\Dcol{\mathbf{G}}(\Mcol{\md(t)})\Bigr]\phi\!\left(\Fcol{\avec}(\Mcol{\md(t)})\odot\left[\zd(t)-\Fcol{\Delta\boldsymbol{\theta}}(\Mcol{\md(t)})\right]\right)+\Fcol{\bvec}(\Mcol{\md(t)})+\Id(t).
    \label{eq:neuromod_all_loci}
\end{equation}
}

{When $\md=0$ and $h(0)=1$, $\mathbf{G}=\mathbf{1}_{N\times N}$, $\Wd=\Wd^{\text{base}}$, $\avec=\mathbf{1}_N$, $\Delta\boldsymbol{\theta}=0$, $\bvec=\bvec^{\text{base}}$, and $\boldsymbol{\tau}=\boldsymbol{\tau}^{\text{base}}$. Equation \eqref{eq:neuromod_all_loci} then yields the unmodulated model dynamics depicted in Equation \eqref{eq:base}.}

{Overall, Equation \eqref{eq:neuromod_all_loci} aims to encompass the space for implementing neuromodulatory drives in ANNs. A majority of published models modulate different subsets of $\{\Scol{\Wd},\Dcol{\boldsymbol{\tau}},\Dcol{\mathbf{G}},\Fcol{\avec},\Fcol{\Delta\boldsymbol{\theta}},\Fcol{\bvec}\}$ with neuromodulation. Because updates across all modulated quantities are governed by the neuromodulatory state vector $\Mcol{\md}$, a single modulator may act at several loci simultaneously, while multiple modulators can converge onto the same locus to express complex multi-neuromodulatory actions. Modulator-specific sensitivity parameters $\{\Scol{\Wd^{(k)}},\Dcol{\boldsymbol{\rho}^{(k)}},\Dcol{\mathbf{K}^{(k)}},\Fcol{\boldsymbol{\mu}^{(k)}},\Fcol{\boldsymbol{\nu}^{(k)}},\Fcol{\boldsymbol{\beta}^{(k)}}\}$ can be constructed from experimental data to serve as biological constraints for ANNs.}

\paragraph{{Co-evolving neuromodulator dynamics}}
{The formulation above specifies how the instantaneous neuromodulatory state $\md(t)$ alters network computation, but neuromodulator concentrations are themselves dynamic variables. Release and co-release of neuromodulators can depend on neural activity, external context, and interactions between neuromodulatory systems, giving rise to reciprocal coupling between network activity and neuromodulatory state \citep{bull2025volume}. A general coarse-grained dynamics for the $K$ neuromodulators can therefore be written as}

{
\begin{equation}
    \boldsymbol{\tau}_{m}\odot\dot{\Mcol{\md}}(t)=-\Mcol{\md}(t) + q\!\left(\mathbf{R}\,\mathbf{r}(t)+\mathbf{Z}\,\Mcol{\md}(t)+\mathbf{c}(t)\right),
    \label{eq:mod_dynamics}
\end{equation}
}

{where $\boldsymbol{\tau}_{m}\in\R_{>0}^{K}$ contains the characteristic timescales of the neuromodulators, $\mathbf{R}\in\R^{K\times N}$ maps neural activity onto neuromodulator release, $\mathbf{Z}\in\R^{K\times K}$ describes effective interactions among neuromodulatory systems, $\mathbf{c}(t)\in\R^K$ represents optional external or contextual influences, and $q$ is a suitable transfer function. The off-diagonal elements $Z_{kl}$ therefore allow neuromodulator $l$ to influence the state or release of neuromodulator $k$, without requiring these terms to correspond to direct chemical interactions. Setting $\mathbf{c}(t)=0$ yields a fully endogenous neuromodulatory system driven by network activity.}

{This distinction reveals several forms of multi-neuromodulator interaction within the same framework. First, multiple neuromodulators can converge on a common locus through sums such as $\sum_k m_k(t)\Wd^{(k)}$. Second, nonlinear mappings such as $h=\exp$ allow these convergent influences to combine multiplicatively. Third, through $\mathbf{Z}$, one neuromodulator can modify the dynamics of another before either signal acts on the network. Consequently, multi-neuromodulatory interactions can arise both at the level of neuromodulator dynamics and at the downstream loci on which those signals converge.}

\paragraph{{Learning rules}}

{It is important to distinguish neuromodulator-driven reconfiguration of effective network parameters from synaptic learning. For example, in the additive connectivity formulation of Equation \eqref{eq:connectivity}, changes in neuromodulator concentrations alone induce a rapid change in effective connectivity according to}

{
\begin{equation}
    \frac{d}{dt}\Scol{\Wd}\bigl(\Mcol{\md(t)}\bigr)=\sum_k\dot{\Mcol{m_k}}(t)\Scol{\Wd^{(k)}}.
    \label{eq:effective_weight_dynamics}
\end{equation}
}

{This change does not itself constitute synaptic plasticity: the parameters $\Wd^{\text{base}}$ and $\Wd^{(k)}$ may remain fixed while the effective connectivity varies with the neuromodulatory state. More generally, for any effective parameterization $\Wd_{\text{eff}}(\md)$, its neuromodulator-driven dynamics follow $\dot{\Wd}_{\text{eff}}=\sum_k (\partial\Wd_{\text{eff}}/\partial m_k)\dot m_k$. This provides a direct link between the dynamics of neuromodulator concentrations and rapid reconfiguration of network computation \citep{bull2025volume}.}

{On a slower timescale, multiple studies have linked neuromodulatory signals to synaptic updates through eligibility traces and three-factor learning rules, whereby local activity-dependent plasticity is gated by subsequent neuromodulatory signals \citep{fremaux_neuromodulated_2016, bellec_solution_2020}. In the present formulation, a generic update of the modulator-specific connectivity profile may be written as}

{
\begin{equation}
    \Delta \Scol{w_{ij}^{(k)}}(t)=-\eta_k\,\Mcol{m_k(t)} \,\ell_i(t)\,e_{ij}(t),
    \label{eq:three_factor}
\end{equation}
}

{where $e_{ij}(t)$ is a local eligibility trace containing pre- and postsynaptic information, $\ell_i(t)$ is a top-down error-, or learning-related factor available to the postsynaptic unit, and $\eta_k$ is the corresponding learning rate. Equation \eqref{eq:three_factor} is intended as a generic multi-neuromodulatory form rather than a unique biological learning rule: the concentration $m_k(t)$ gates the update associated with the sensitivity profile of neuromodulator $k$. This construction is closely related to context-gated learning rules in which neuromodulator concentrations multiply otherwise local synaptic-update terms \citep{bull2025volume}.}

{These frameworks naturally accommodate modulator-specific gating, while neuromodulatory changes in neuronal gain, time constants, and other intrinsic properties further reshape the activity statistics from which synaptic eligibility traces are generated. With multiple neuromodulators, distinct concentrations and receptor-specific sensitivities can therefore scale different components of plasticity, defining multiple directions of adjustment in parameter space. When modulator-specific sensitivity profiles are constrained to a shared low-dimensional basis, or collectively span only a small subspace of the full parameter space, the accessible update directions are correspondingly restricted, biasing learning toward a structured and lower-dimensional family of parameter updates.}

\paragraph{{Evaluating functional equivalence}}
{Adapting brain-inspired computational principles across spatio-temporal scales demands rigorous testing of their functional overlaps and possible redundancy, as seemingly distinct mechanisms can lead to dynamics-level degeneracy. In addition, evaluating the functional equivalence of proposed implementations helps identify the principal axes of neuromodulatory control and determine whether ostensibly distinct mechanisms provide independent or redundant forms of network adaptation.}

{An important distinction is therefore between functional equivalence and mechanistic correspondence. Generic machine-learning mechanisms such as feature-wise modulation, hypernetworks, low-rank adaptation (LoRA), or meta-learned control signals can produce context-dependent changes in network computation that are functionally similar to neuromodulation \citep{perez_film_2018, hu2022lora, ha_hypernetworks_2017, vecoven_introducing_2020}. However, their control variables and adaptation parameters are generally optimized for task performance rather than explicitly constrained to correspond to neuromodulatory mechanisms or specific loci of action. In contrast, our formulation explicitly separates the neuromodulatory state $\md(t)$ from modulator-specific sensitivity profiles acting on connectivity, intrinsic dynamics, state-to-output transformations, and plasticity. The additional matrices $\mathbf{R}$ and $\mathbf{Z}$ separately specify how neural activity and interactions among neuromodulatory systems determine the evolution of the neuromodulatory state itself. While this decomposition does not inherently offer more expressive power than standard machine-learning parameterizations, it can improve mechanistic interpretability by assigning distinct and experimentally grounded computational roles to neuromodulatory variables and their loci of action \citep{bull2025volume, liu_cell-typespecific_2021}. Corresponding in silico manipulations of neuromodulator concentrations, sensitivity profiles, interactions, or loci of action can therefore be related more directly to experimental perturbations. The principal advantage of this formulation is thus its interpretability and falsifiability, allowing functionally similar computational mechanisms to be distinguished according to their consistency with experimentally observed neuromodulatory dynamics.}

\subsection{Neuromodulation-inspired components across scales}
Emulating the intricate interplay of morphology, neuronal dynamics, and neuromodulatory processes promotes learning in complex environments, potentially advancing the capabilities of ANNs. Here, we leverage multiple network scales on which neuromodulatory signals regulate learning, based on their spatio-temporal complexity in the brain. 

\subsection{Subcellular and neuronal level} 
The structural complexity of neurons is instrumental in information processing. Dendritic heterogeneity, which refers to the variation in dendritic branching and spine density, allows neurons to integrate diverse inputs effectively. Mimicking this in DNNs involves designing models that can adaptively modify their connectivity patterns, enabling more nuanced feature extraction and representation learning. 

Neuronal dynamics, including spiking behavior and receptor modeling, are essential for temporal information processing and synaptic plasticity. Incorporating spiking mechanisms into DNNs can enhance their ability to handle sequential and time-dependent data. Additionally, modeling neuronal heterogeneity, where neurons exhibit diverse response patterns, can lead to more versatile network behaviors.

At the subcellular and neuronal level, structural and functional complexity can be incorporated through dendritic compartments and learnable biases in ANNs. In SNNs, these properties can be realized through different spiking behaviors.

\paragraph{Structural diversity} Incorporating dendritic architectures enables resilient continual learning \citep{acharya_dendritic_2022, pagkalos_leveraging_2024} and offers a plausible explanation for backpropagation signals \citep{greedy2022single, payeur_burst-dependent_2021, sacramento2018dendritic}. \citep{iyer_avoiding_2022} proposed an architecture featuring context-driven dendritic layers and capable of learning multiple tasks with minimal forgetting. Similarly, multi-task learning can be achieved through NMDA-driven dendritic modulation in a self-supervised biophysical model, where task-dependent modulations are applied to individual neurons \citep{wybo_nmda-driven_2023}. Incorporating temporal diversity also enables dendrites to function as temporal gates, leading to multi-timescale learning \citep{zheng_temporal_2024}. 

Combined with multi-compartmental morphology, neuromodulation brings a new dimension for tuning neuronal responses. Neuromodulation-inspired mechanisms allow contextual cues to shape dendritic processing, while compartmental models help highlight the role of neuromodulators in promoting dendritic and spine structural plasticity. Dendritic properties can be modified through parameters including dendrite length, diameter, and branching. 

\paragraph{Functional diversity} Heterogeneous neuronal dynamics is fundamental to biological systems \citep{fan_towards_2025, izhikevich_which_2004, kanari_objective_2019, markram_interneurons_2004, rich_loss_2022} and is frequently overlooked in DNNs. A study on ANNs leverages neuronal bias for multi-task learning through backward transfer, underscoring the importance of functional heterogeneity \citep{williams2025expressivity}. In SNNs, neuronal timescale heterogeneity in leaky integrate-and-fire (LIF) neurons enhances robustness \citep{perez-nieves_neural_2021}, while a temporal hierarchy within the network leads to higher performance \citep{moro2026temporal}. \citep{habashy_adapting_2024} employed evolutionary algorithms to investigate bursting parameter heterogeneity, allowing the network to solve spatio-temporal tasks. Theoretical work with heterogeneous SNNs of Izhikevich neurons demonstrated that adaptive network computations were achieved at the spike level \citep{gast_neural_2024}, offering a mechanism by which neuromodulatory effects on neuronal dynamics could be implemented. 

Given the computational costs associated with simulating subcellular and neuronal-level neuromodulatory processes (e.g., ion channel parameters), higher-level abstraction is possible using a process similar to simplifying biophysical models and representing the overall properties, such as adaptive firing threshold, which shifts the gain function at the population level \citep{shine_computational_2021}. Furthermore, axonal and dendritic propagation delays, a mechanism that is often overlooked in computational studies, can contribute to the emergence of connectivity patterns. Neuromodulators can alter the excitability of axons, and therefore, the temporal fidelity. Introducing neuronal-level heterogeneity can help link these neuron-level mechanisms with network-level dynamics.

\paragraph{Receptor dynamics} ANNs primarily emulate rapid feedforward information flow, representing short-term plasticity mechanisms analogous to ionotropic receptor dynamics. However, the slower and more complex processes governed by metabotropic receptors, which play a critical role in higher-order cognitive representations, are often unexplored \citep{hasselmo_unexplored_2021}. Although \citep{fremaux_neuromodulated_2016} incorporated aspects of metaplasticity through neuromodulated learning rules, broader metabotropic effects -- such as modulation of neuronal excitability, gain or gating -- were still missing. Modeling these processes is challenging, as metabotropic signaling depends on intracellular cascades that are computationally expensive to simulate in artificial systems.

Recurrent architectures such as gated recurrent units (GRUs) and long short-term memory networks (LSTMs) capture longer temporal dependencies and offer partial abstractions of gating. Neuromodulatory signals have been functionally approximated by mechanisms analogous to the forget gate in LSTM cells \citep{costacurta_structured_2024}. Similarly, training weights and biases independently raises the possibility of interpreting bias modulation as a proxy for neuromodulation-driven regulation \citep{williams2025expressivity}. Nevertheless, achieving precise control over single-neuron dynamics requires models that explicitly represent spike-based computations. SNNs provide such a framework and offer a substrate for studying metabotropic-like modulation on excitability and plasticity \citep{rodriguez-garcia_enhancing_2024}.

\subsection{Circuitry level}
The circuitry level takes into account both structural elements such as neuronal connectivity and population-level diversity of neurons (micro-circuitry), as well as the emergence of neuronal populations into functionally specific groups and the interconnections across these groups (meso-circuitry). Unlike ANNs, in the biological brain, connectivity between neurons features a sparse pattern and is determined by multiple factors. Such connectivity facilitates energy efficiency, evolves with development, and underlies plasticity and learning. 

\paragraph{Connectivity} The microcircuit structure of biological neural networks varies across brain regions. Bio-inspired DNNs attempt to mimic this complexity by imposing constraints on synaptic weight plasticity, adjusting connectivity features such as sparsity and connection probability \citep{lachi_stochastic_2024, perez-nieves_neural_2021, yang_towards_2021}. Neuronal connectivity is shaped by factors such as the genetic type of neurons and the distance between neurons \citep{billeh_systematic_2020, markram_reconstruction_2015, stoeckl_structure_2021}. Compared to DNNs with fully-connected layers, the connection probability between neurons in the brain is low, leading to sparse connectivity \citep{mocanu_scalable_2018}, which can be introduced in DNNs by adding a non-trainable sparse matrix to define the network connectivity \citep{yang_towards_2021}. Additionally, a recent approach explored stochastic wiring by incorporating connection probabilities, highlighting that randomness in connectivity might be an evolutionarily developed feature in biological organisms \citep{lachi_stochastic_2024, perez-nieves_neural_2021}.

Neuromodulation can regulate the global connection profile of neural networks. Meanwhile, network connectivity shapes network dynamics changes caused by the neuromodulatory tone \citep{rich_effects_2020}. Given the multitude of pre- and post-synaptic processes neuromodulators affect, they not only participate in regulating the probability of connection but its strength. Neuromodulation plays an important role in modifying circuit-level connectivity through both direct and indirect mechanisms: For example, they contribute to the formation and elimination of synapses (direct mechanism), while also altering the excitability of neurons (indirect mechanism) \citep{nadim_neuromodulation_2014}. Moreover, one synapse may be under the influence of multiple neuromodulators, and the combined effects may not be additive and depend on the network state \citep{koh_two_2003, nadim_neuromodulation_2014}.

\paragraph{Excitation and inhibition} Dale’s principle \citep{dale_pharmacology_1935, eccles_cholinergic_1954} led to the introduction of excitatory and inhibitory neuronal populations into ANNs and has been challenged by the finding that neurons can release more than one neurotransmitter \citep{tritsch_mechanisms_2016, vaaga_dual-transmitter_2014}. In practice, this imposes a constraint on the sign of synaptic weight in the network, meaning that excitatory neurons are restricted to facilitating positive signal transmission and inhibitory neurons are limited to negative signal transmission \citep{yang_towards_2021}. A few DNN and recurrent neural network (RNN) architectures have incorporated these features, often adopting the 80:20 excitation:inhibition ratio \citep{cornford_learning_2020, kao_considerations_2019, li2023learning, song_training_2016}. A study on SNNs highlighted the significance of this specific ratio, showing that it leads networks to reliably train at low activity levels and in noisy environments, underscoring its practicality \citep{kilgore2025biologically}. However, while bio-inspired, this constraint may limit the learning and performance of DNNs as the available parameter space is reduced \citep{cornford_learning_2020, kao_considerations_2019, li2023learning}. The addition of neuromodulatory signals at this level could enable switching between excitatory and inhibitory weights, allowing networks to better adapt to specific tasks and support multi-task learning by preserving weight signs across sequential tasks.

\subsection{Network level}
Neuromodulatory systems play a crucial role in shaping large-scale brain network dynamics. These systems extend their influence beyond localized circuits, modulating brain-wide activity patterns and facilitating the coordination of diverse cognitive functions \citep{marder_neuromodulation_2012, mei_effects_2023}. At this broader systems level, understanding the impact of global neuromodulatory signals on large-scale networks offers valuable insights for the design of ANNs.

\paragraph{Network topology} Network neuroscience provides tools for unveiling the implications of brain structures and their emerging properties, as well as an analytical framework for studying the neuromodulatory systems. Network neuroscience research employs graph theory and treats the brain as an interconnected network of nodes (regions) and edges (connections). Key measures such as modularity, integration, and participation coefficient offer insights into the organization and efficiency of these networks. Neuromodulatory systems dynamically adjust connections and promote efficient communications across and within brain regions, and play a vital role in maintaining the balance between network segregation and integration, which is essential for robustness, efficiency and adaptability \cite{shine_neuromodulatory_2019}.

Open source brain atlas and data sharing have allowed a closer examination of the brain's network, offering a data-intensive view of its specialized structural and functional modules responsible for perceptual, cognitive and motivational tasks \citep{hansen_mapping_2022}. Network topologies identified in biological neural networks have been used to construct ANNs with reduced number of parameters but no performance decline \citep{goulas_bio-instantiated_2021, mocanu_scalable_2018}. Moreover, topologies derived from connectome data are shown to promote efficient reinforcement learning when incorporated into SNNs \citep{wang_brain_2024}. 

Determining the modularity of brain networks and superimposing it with the spatial domains of neuromodulation, then probing its convergence and divergence across functions and brain states, may serve as a guide to create specialized network topologies for different tasks. In the meantime, identifying nodes and clusters shared between tasks may shed light on a unifying view of centralized processing across task domains.

\paragraph{Hyperparameters and activation functions}
Neuromodulation-inspired mechanisms enable ANNs to adjust hyperparameters and activation functions, thus regulating network dynamics in response to changing environments, task demands and cognitive/behavioral states \citep{brna2019uncertainty, mei_effects_2023, vecoven_introducing_2020, wilson_neuromodulated_2018}. Such adaptability is crucial for intelligent systems capable of operating autonomously in real-world environments where conditions and requirements frequently shift. The capacity for adaptive reconfiguration at the network scale is critical for improving the resilience of ANNs, enhancing their ability to withstand disruptions or noise while maintaining stable performance. 

\paragraph{Global multi-neuromodulatory interactions} Inspired by \citet{doya_metalearning_2002}, deep reinforcement learning (DRL) offers a structured approach for interpreting neuromodulatory actions. In DRL, neuromodulatory functions can be easily mapped to model hyperparameters: DA for reward prediction through temporal difference (TD) learning, 5-HT controls the influence of short- and long-term rewards, NA modulates the randomness of action selection through a Softmax policy, and ACh affects the learning rate \citep{doya_metalearning_2002, krichmar_neuromodulatory_2008, mei_informing_2022}. These depictions have been widely used to support lifelong RL in artificial agents \citep{ben-iwhiwhu_context_2022, lee_lifelong_2024, mei_informing_2022}. However, this one-to-one mapping between neuromodulatory signals and their functional role through single hyperparameters may be an oversimplification. 

Hence, understanding the synergistic and balancing interactions among neuromodulators is crucial for the design of more sophisticated models that replicate human-like decision-making and problem-solving abilities. In biological systems, neuromodulatory systems do not operate in isolation; they interact continuously in a state- and context-dependent manner. These interactions fine-tune various neurobiological processes essential for adaptive behavior \citep{avery_neuromodulatory_2017, brzosko_neuromodulation_2019} and for acquiring new information with minimal interference \citep{bradfield_thalamostriatal_2013, matityahu_acetylcholine_2023}. 

In ANNs, multi-neuromodulatory mechanisms can be introduced through \textbf{(i)} regulation and refinement of neuromodulatory drives through other neuromodulators, \textbf{(ii)} spatial and temporal correlations of neuromodulatory drives, and \textbf{(iii)} task-specific behavior of localized and global neuromodulation. 

\section{Multi-neuromodulatory dynamics in ANNs: a conceptual model}
\begin{figure}[ht]
\centering
\includegraphics[width=1\textwidth]{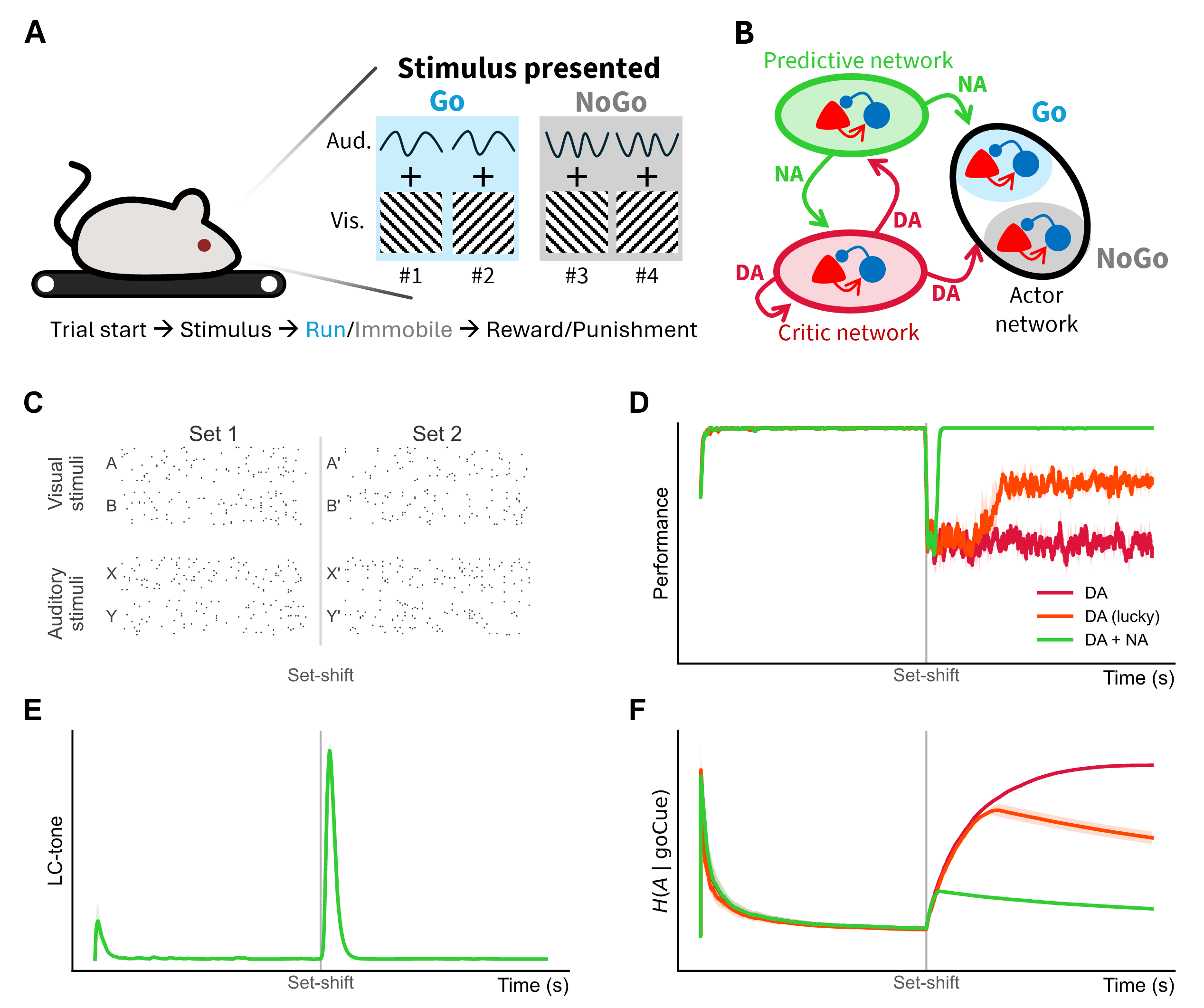}
\caption{\textbf{Contingency adaptation in spiking neural networks during a Go--No-Go task.} \textbf{A)} Experimental paradigm. \textbf{B)} Neuromodulatory actor--critic architecture, comprising a predictive network and a critic network, with DA and NA signals regulating learning and gain. \textbf{C)} Input structure before and after the set shift. Two visual stimuli and two auditory stimuli are encoded as sparse population activity. \textbf{D)} Task performance over time for DA-only learning (red), DA-only ``lucky'' trials that successfully adapt by chance (orange), and DA + NA co-modulated learning (green). The vertical line represents the set shift. \textbf{E)} The LC-tone transient at the set shift. \textbf{F)} Conditioned action entropy given the go cue, $H(A \mid \mathrm{goCue})$, quantifying exploratory behavior. DA: dopamine; NA: noradrenaline; LC: locus coeruleus.}
\label{fig:F3}
\end{figure}

To illustrate the effects of multi-neuromodulatory dynamics, we present a conceptual study using a \textbf{reward-driven, extra-dimensional/intra-dimensional set-shifting Go/No-Go task} \citep{dias_dissociation_1996, konishi_transient_1998, robbins_shifting_2007}; Figure \ref{fig:F3}A. In this paradigm, the agent must respond to pairings of two sets of input patterns representing distinct sensory modalities (e.g., visual (A, B) and auditory (X, Y) stimuli; Figure \ref{fig:F3}C) and must learn, through trial and error, which stimulus predicts a reward. The agent’s objective is to acquire and maintain an appropriate response policy under a given contingency. Crucially, midway through the experiment, the stimulus–reward contingency is altered, such that the previously rewarded “Go” cue no longer applies, forcing the agent to rapidly adapt its behavior to a new contingency.

We examine how DA-like signals support reward-driven plasticity under stable contingencies, while NA-like signals enable rapid reconfiguration and exploration following contingency shifts -- capturing key aspects of biological learning flexibility that emerge only when both neuromodulators act in concert. Motivated by reinforcement learning strategies, we consider an actor–critic architecture \citep{chung_reinforcement_2020, dayan_theoretical_2001, sutton_reinforcement_2020} augmented by a predictive coding sub-module for detecting unexpected changes in the environment \citep{barry_fast_2024}, thereby facilitating flexibility when reward contingencies change \citep{wainstein_2025, munn_neuronal_2023, rodriguezgarcia2025noradrenergic} -- see Figure \ref{fig:F3}B. {Within the formulation introduced above, the predictive network implements the neuromodulator-dynamics layer, whereas the resulting DA- and NA-like signals operate at distinct loci: DA primarily at the learning-rule level and NA at the level of neuronal state-to-output transformations.}

\begin{itemize}
    \item The \textbf{actor network} is responsible for the agent’s decision-making. We use two groups of neurons (one for each action 'Go' (blue) or 'No-Go' (grey)), each containing excitatory and inhibitory units. The network selects the action depending on the group that has a higher net firing rate, i.e., the average firing rate of the excitatory neurons minus that of the inhibitory neurons.

    \item The \textbf{critic network}, composed of excitatory and inhibitory neurons, estimates the value of the expected reward in the current state by its net firing rate. The TD error is computed by comparing the estimated value with the actual reward received during the trial \citep{schultz_neural_1997}. The RPE represents the DA activity of the network and acts as a third factor that modifies synaptic plasticity, following the R-STDP learning rule \citep{chung_reinforcement_2020}. Specifically, DA activity adjusts the reward expectancy based on current contingency.
    {This modulation can be expressed as $\Delta w_{ij}(t) = \eta\,m_{\mathrm{DA}}(t)\,e_{ij}(t)$, where $e_{ij}(t)$ is the synaptic eligibility trace and $m_{\mathrm{DA}}(t)$ is the RPE-driven DA-like signal acting at the plasticity layer of the actor--critic network.}

    \item The \textbf{predictive network} detects changes in task contingencies using principles inspired by predictive coding. It continuously compares incoming stimuli with predictions based on previous experiences to identify discrepancies or unexpected changes \citep{barry_fast_2024}. When the network detects such a change in contingency -- that is, when the expected relationship between stimuli and outcomes alters -- it responds by releasing a neuromodulatory signal modeled after NA. By emulating the LC's dynamics, the signal is appropriately timed to influence the network's processing during set-shifting periods (Figure \ref{fig:F3}E). {Following the general formulation introduced above, the dynamics of this NA-like signal are described by $\tau_{\mathrm{NA}}\dot{m}_{\mathrm{NA}}(t)=m_{\mathrm{NA}}^{0}-m_{\mathrm{NA}}(t)+\omega\epsilon(t)$, where $m_{\mathrm{NA}}^{0}$ denotes the tonic baseline and $\epsilon(t)$ the signal driving phasic NA release \citep{wainstein_2025}. In this conceptual implementation, $\epsilon(t)$ is an information-based signal representing the reduction in uncertainty about the selected action $A$ when it is known that the action was correct, estimated over a sliding window $\mathcal{W}_t$ of previous episodes.} The NA-like signal targets excitatory neurons in both the actor and critic networks, promoting synaptic plasticity by inducing correlated bursting activities in these neurons. {At the level of the proposed formulation, this NA-like mechanism acts primarily on the intrinsic and state-to-output transformation layer of excitatory units in the actor and critic networks, transiently modulating their excitability though gain (Equation \ref{eq:gain_a}).} As in \citet{munn_neuronal_2023, rodriguezgarcia2025gainlayer5}, such bursting elevates the arousal state, flattening the energy landscape and enhancing network flexibility \citep{aston-jones_integrative_2005}. Hence, this heightened arousal facilitates the reconfiguration of the network, allowing it to adapt more effectively to new contingencies by promoting the exploration of alternative actions \citep{bouret_network_2005, doya_metalearning_2002, usher_role_1999}. Intuitively, the NA-inspired component will dynamically flatten the loss landscape, enhancing a transient exploration between attractors \citep{wainstein_2025, rodriguezgarcia2025noradrenergic}.
\end{itemize}

\subsection{Evaluation and interpretation}
Model performance can first be evaluated under stable contingencies, assessing how the DA-like signal modulates reward-driven plasticity through an R-STDP rule. In this setting, learning efficacy is quantified by the ability of the actor-critic network to acquire and exploit rewarded actions, with the TD error shaping synaptic updates in the critic network, consistent with established neuromodulated learning frameworks \citep{chung_reinforcement_2020, florian_reinforcement_2007, fremaux_neuromodulated_2016}. Following a contingency shift, evaluation focuses on whether the NA-like signal promotes flexible network reconfiguration after changes are detected by the predictive module. This flexibility is quantified by the randomness or uncertainty of action selection, where increased exploration corresponds to higher uncertainty and increased exploitation to lower uncertainty, measured using the Shannon entropy of actions conditioned on correct responses:
\begin{equation*}
H(A \mid \mathrm{goCue}) =-\sum_i P(a_i \mid \mathrm{goCue}) \log_2 P(a_i \mid \mathrm{goCue}).
\end{equation*}
where $P(a_i \mid \mathrm{goCue})$ denotes the probability of selecting action $a_i$ when the Go cue is presented. When a set-shift occurs (i.e., a change in contingency), the system no longer knows the correct responses for the new task. Therefore, the entropy increases again as the agent explores the new contingency (Figure \ref{fig:F3}F). However, since the network’s weights were optimized for the previous contingency, the timescale for adapting to the new one depends on how quickly the system can reconfigure its weights. This can result in prolonged optimization times converging toward suboptimal solutions or, in most cases, failure to optimize altogether (red and orange lines, Figures \ref{fig:F3}D and F). Introducing NA (Figure \ref{fig:F3}E) into the system facilitates weight reconfiguration: NA promotes exploration by increasing the entropy temporarily, enabling the system to rapidly sample and evaluate new actions. This accelerates the discovery of the new contingency, leading to faster adaptation compared to relying solely on RL mechanisms (green lines, Figures \ref{fig:F3}D and E).

This conceptual model illustrates how neuromodulation-aware spiking neural networks can integrate stable reward-based learning with rapid adaptation to changing contingencies. DA-like signals guide learning under stable conditions through third-factor plasticity \citep{chung_reinforcement_2020, izhikevich_solving_2007}, while NA-like signals transiently increase network flexibility following detected changes, promoting exploration and facilitating dimensionality shifts \citep{rodriguezgarcia2025noradrenergic, wainstein_2025}. The resulting dynamics provide mechanistic intuition for how exploration and exploitation trade offs emerge from interactions between neuromodulatory systems, enabling continual adaptation in non stationary environments. They also point to key challenges for future work, including the experimental measurement of conditioned action entropy to evaluate these intuitions, the development of accurate SNN implementations, and the theoretical study of richer neuromodulatory interactions.

\section{Discussion and outlook}
{In this Perspective, we have argued that neuromodulation is best understood and implemented not as a set of isolated, single-modulator-driven processes, but as a highly coordinated control layer over learning and adaptive behavior. Such flexible, state-dependent control is facilitated by multiple chemically distinct signals that interact across timescales and can be implemented in ANNs during both forward computation and learning. The specific contribution of the multi-neuromodulatory perspective does not lie in any single locus of implementation, but in the contextualization of individual neuromodulator effects within a framework where multiple neuromodulators are concurrently in action, providing space for concretely realizing the spatial (across subneuronal domains, neuron types, circuits and pathways) and temporal neuromodulation of learning (across timescales of immediate adaptation, consolidation, and long-term integration).}
{\subsection{What a multi-neuromodulator framing contributes beyond existing ML models}
The framing differs from existing ML approaches to continual learning. First, we identify three loci of neuromodulatory actions where structural, dynamical and functional modulation can be concurrently achieved, allowing neuromodulatory mechanisms to be used in the forward pass. This way, complex interactions between neuromodulators (e.g., opponent, convergent and modulatory) can also be represented. Published research already demonstrates the computational value of representing neuromodulation as a multidimensional state. In the e-nmRNN of \citet{bull2025volume}, distinct neuromodulatory components independently modulate effective connectivity, such that a low-dimensional neuromodulatory state dynamically reconfigures a high-dimensional network. In this parameterization, different modulators can provide distinct directions of context-dependent modulation, and increasing the number of non-redundant modulatory components can therefore enlarge the subspace of network configurations that can be represented. Our formulation of $\md(t)$ provides a substrate for an analogous factorization across multiple loci of neuromodulatory action, while allowing the identity, dynamics, and interactions of individual neuromodulatory systems to be informed by experimental findings.}

{A second distinction from current continual learning approaches is that $\md(t)$ need not be specified by an explicit task label (task-agnostic). Instead, its dynamics can depend on signals such as reward prediction error, uncertainty, task relevance, and contextual evidence. This provides a route to context-dependent adaptation in which changes in the modulatory state are inferred from ongoing signals rather than triggered by externally supplied task boundaries.}
\subsection{Exploring multi-neuromodulatory mechanisms}
A useful framing is to view neuromodulatory control through the lens of dynamical systems and attractor landscapes \citep{amit_modeling_1989, freedman_experience-dependent_2006, li_robust_2016, mante_context-dependent_2013, romo_neuronal_1999}. In this view, neuromodulators can stabilize task-relevant states, reshape their basins of attraction, or trigger transitions when contingencies change, thereby balancing plasticity and stability \citep{munn_neuronal_2023, shine_neuromodulatory_2023, wainstein_2025, rodriguezgarcia2025gainlayer5}. This perspective naturally accommodates cooperation and antagonism between neuromodulators and suggests algorithmic motifs for continual learning, where different signals gate learning, exploration, and consolidation under task shifting.

Multi-neuromodulator dynamics flexibly promote multiple learning paradigms such as transfer-learning, meta-learning, and incremental learning, depending on the particular constraints imposed on the agent. However, there are a number of challenges in the study of neuromodulation and in neuromodulation-aware ANNs. A prominent example is neuromodulatory projections to specific subtypes of neurons and their region-dependent, projection-specific effects. Consequently, the use of modular architectures in neuromodulation-aware ANNs may help implement modulatory projections to specific subsets of units. Nevertheless, it remains to be studied how such modular architectures in ANNs could be designed in a principled manner \citep{rodriguez-garcia_enhancing_2024, yang_towards_2021}. More importantly, the complexity of neuromodulatory systems and neural circuitry in the biological brain (e.g., neuronal heterogeneity, tonic and phasic firing, and multi-neuromodulatory interactions) serves as a foundation of the distinct neuromodulatory dynamics, and the complex interplay between neuromodulators and their receptors. Despite recent developments in experimental neuroscience techniques, several challenges persist (Box \ref{box:B3}). The multi-scale effects of neuromodulators, the limited resolution of pharmacological and neurogenetic tools, and the prevalent co-release of neuromodulators and neurotransmitters complicate the study of neuromodulatory mechanisms, highlighting the need for the development of new technical tools.

Computational models provide an important bridge across spatial and temporal scales, linking synaptic and cellular mechanisms to population dynamics, behavior, and cognition. Biophysical models connect synaptic processes with spiking activity, while network models relate collective neural dynamics to learning and adaptive behavior. Incorporating multi-scale neuromodulatory dynamics into such models allows for systematic investigation of how neuromodulation shapes plasticity, network organization, and behavioral outcomes, facilitating the generation and testing of experimentally grounded hypotheses. Seminal examples include cerebellar learning frameworks \citep{albus_theory_1971, ito_long-lasting_1982, marr_theory_1969} and reinforcement learning models of DA signaling \citep{schultz_neural_1997, sutton_reinforcement_2020}. Going beyond basic neuroscience research, computational models also provide powerful tools for studying brain disorders \citep{carannante_impact_2024, lanillos_review_2020, lindroos_basal_2018, pavlides_computational_2015, verzelli_editorial_2024}, revealing how disruptions in neural circuits can cause physiological and cognitive dysfunctions, informing the development of therapeutic strategies for neurological and psychiatric diseases \citep{montague_computational_2012, frank_by_2004}.

\begin{featurebox}
\caption{\small{Methodological challenges in experimental studies of the neuromodulatory system}}
\label{box:B3}
\footnotesize{
    \textit{\textbf{Pharmacological manipulations:}} Many studies on neuromodulator interactions rely on pharmacological manipulations of receptor subtypes at a systems level, limiting insights into local endogenous release and interactions of neuromodulators. For experiments conducted under awake conditions, key challenges include limited spatiotemporal resolution, possible off-target effects, varying dose-response relationships across subjects, compensatory mechanisms, and imperfect receptor specificity \citep{grossman_neuromodulation_2022}. Furthermore, pharmacological studies conducted in strictly controlled environments often fail to capture the context-, state- and task-dependent effects of neuromodulators. 

    \textit{\textbf{Neurogenetic tools:}} G-protein-coupled receptor activation-based (GRAB) sensors for neuromodulators such as DA and NA have overcome the limitations of pharmacological manipulations \citep{feng_genetically_2019, sun_next-generation_2020}. Although they have advanced real-time detection and measurement of neuromodulator actions \citep{doya_serotonergic_2021}, their use is still limited due to insufficient spatial and temporal resolutions. Similarly, transgenic animal models (e.g., Cre-driver lines) offer powerful tools for targeting specific neuronal populations. However, off-target expression of marker genes in unintended cell types or brain regions can affect the reliability and interpretation of experimental results. In \citep{ren_anatomically_2018}, for some transgenic mouse lines, Cre protein expression, which is supposed to target only 5-HT-producing neurons, is not restricted to these neurons and leads to recombination in non-5-HT neurons, potentially confounding the identification and interpretation of 5-HT-related functions.

    \textit{\textbf{Precise targeting of neuromodulators:}} Given the diverse connections to and from neuromodulator-releasing cells \citep{watabe-uchida_whole-brain_2012}, it is crucial to examine the physiological and behavioral effects of neuromodulation in a projection- and neuron-type-specific manner. However, several issues must be addressed: \textbf{(i)} Quantifying neuromodulator levels and their receptor activities in some brain regions is challenging given their low concentrations and rapid release and uptake. \textbf{(ii)} Selectively manipulating neuromodulatory pathways is difficult due to the extensive and overlapping projection patterns of neuromodulatory systems. \textbf{(iii)} Capturing and delineating neuromodulatory dynamics and effects across timescales, ranging from rapid changes in neuronal excitability to long-term alterations in gene expression, requires sophisticated experimental designs and analytical methods.

    \textit{\textbf{Disentangling the co-release of neuromodulators:}} The complexity of neurotransmitter co-release makes it difficult to elucidate the effects of individual neuromodulators. For instance, cholinergic interneurons (CINs) in the striatum, which are central to ACh signaling, are known to co-release glutamate and gamma-aminobutyric acid (GABA). This complicates the interpretation of experimental results. \citet{matityahu_acetylcholine_2023} investigated the local effects of CIN activity on DA in the striatum and suggested that depending on the context and experimental condition, CINs can enhance or suppress DA release. Another example is midbrain DA neurons, which can co-release glutamate from their axonal terminals, with VGLUT2 playing a crucial role in this process \citep{dal_bo_dopamine_2004, stuber_reward-predictive_2008, sulzer_dopamine_1998}. 
}
\end{featurebox}

\subsection{Implementing and interpreting multi-neuromodulatory mechanisms in ANNs}
\subsubsection{Challenges and avenues}

The translation of neuromodulatory principles faces two bottlenecks. First, multi-scale neuromodulation expands the parameter space, while biological parameters rarely map cleanly onto deep learning hyperparameters, complicating convergence and reproducibility between frameworks \citep{mei_effects_2023}. Second, interpreting the effects of neuromodulation-inspired rules is challenging as similar behaviors can arise from different modulatory configurations \citep{tononi_measures_1999}. 

To address interpretability issues and the black-box nature of DNNs, explainable AI approaches can be extended to quantify how neuromodulatory signals shape network activations, learning dynamics, and decisions. Attention-based and graph-based methods, combined with perturbations of specific modulatory components, may help clarify whether and how neuromodulation shapes computation. Complementing these approaches with experimentally-grounded tasks, benchmarks and analyses can improve understanding of model performance and associated mechanisms. Together with neuromodulated deep learning architectures, these tools provide a foundation for building biologically-informed yet interpretable models that capture the spatio-temporal complexity of neuromodulation.

\subsubsection{Interdisciplinary collaborations and community-driven efforts}
The computational cost and scalability of biologically detailed models remain significant obstacles for deploying neuromodulation-aware systems in large-scale applications. Although the application of neuromodulatory principles to ANNs has been previously proposed, gaps in our knowledge and translational efforts still hinder its development. Progress toward neuro-inspired adaptive learning systems, which is dependent on the integration of experimental insights with principled computational abstractions, requires collaborations between neuroscience and AI experts \citep{kudithipudi_biological_2022}.

A way forward is the development of community-driven resources for neuromodulation-aware ANNs, including shared datasets, benchmarks, models, and tools. Such platforms could host experimental and simulated multi-scale data alongside biologically grounded tasks that enable comparison across methods. Coordinated efforts to pool data across laboratories and species \citep{kelberman_diversity_2024}, together with open theoretical and computational frameworks \citep{wheeler_hippocampomeorg_2015, ramaswamy_neocortical_2015}, will accelerate the integration of biological principles into ANNs. Additionally, interdisciplinary venues focusing on neuromodulation in learning, e.g., \textit{Neuromodulation of Adaptive Learning: Theoretical Lessons Learned From Invertebrate and Vertebrate Brains, 2024 (TP24NM)} organized at the Okinawa Institute of Science and Technology (OIST), facilitate the exchange of experimental and computational methods on neuromodulatory principles in biological neural networks and their artificial counterparts. Through cross-disciplinary dialogue, these initiatives can support the development of more robust, interpretable, and adaptive artificial systems inspired by neuromodulated learning in the biological brain.



\bibliography{elife-sample}

\end{document}